\documentclass[11pt]{article}

\usepackage{latexsym}
\usepackage{amssymb}
\usepackage{amsmath}
\usepackage{mathrsfs}
\usepackage{amsfonts}
\usepackage{amsthm}
\usepackage{epsfig}
\usepackage{graphicx,psfrag,subfigure} 

\newcommand{\Xp}{\mbox{\boldmath $X$}}

\newcommand{\xp}{\mbox{\boldmath $x$}}
\newcommand{\thetap}{\mbox{\boldmath $\theta$}}
\newcommand{\thetaps}{\mbox{\scriptsize\boldmath $\theta$}}

\newcommand{\Yp}{\mbox{\boldmath $Y$}}

\newcommand{\eqa}{\mbox{$ \;\stackrel{(a)}{=}\; $}}
\newcommand{\eqb}{\mbox{$ \;\stackrel{(b)}{=}\; $}}
\newcommand{\eqc}{\mbox{$ \;\stackrel{(c)}{=}\; $}}

\newcommand{\PR}{\mbox{\rm Pr}}

\def\be{\begin{eqnarray}}
\def\ee{\end{eqnarray}}
\def\ben{\begin{eqnarray*}}
\def\een{\end{eqnarray*}}

\def\elabel#1{\label{e:#1}}

\def\sq{$\Box$}

\def\qed{\ifmmode\sq\else{\unskip\nobreak\hfil
\penalty50\hskip1em\null\nobreak\hfil\sq
\parfillskip=0pt\finalhyphendemerits=0\endgraf}\fi\par\medbreak}

\newsavebox{\junk}
\savebox{\junk}[1.6mm]{\hbox{$|\!|\!|$}}

\def\til={{\widetilde =}}

 \def\eq#1/{(\ref{#1})}

\newtheorem{theorem}{Theorem}[section]

\newtheorem{proposition}[theorem]{Proposition}
\newtheorem{lemma}[theorem]{Lemma}

\def\eq#1/{(\ref{e:#1})}

\newcommand{\beqn}[1]{\notes{#1}%
\begin{eqnarray} \elabel{#1}}

\newcommand{\eeqn}{\end{eqnarray} }

\newcommand{\beq}[1]{\notes{#1}%
\begin{equation}\elabel{#1}}

\newcommand{\eeq}{\end{equation}} 

\def\bdes{\begin{description}}
\def\edes{\end{description}}

\def\notes#1{}

\newcounter{tasks}
\newenvironment{MyEnumerate}%
{\begin{quote}%
\vspace*{-0.05in}\begin{list}{\hspace{-.75cm}
		\hbox{\textrm{\rm ({\bf\roman{tasks}})}\ }}
		{\usecounter{tasks}%
        \setlength{\labelsep}{0pt}
        \setlength{\leftmargin}{10pt}
        \setlength{\rightmargin}{0pt}
        \setlength{\labelwidth}{0pt}
        \setlength{\listparindent}{0pt}}}%
{\end{list}\end{quote}}

\title{
 	Estimating the Entropy of Binary Time Series:\\
	Methodology, Some Theory and a Simulation Study
}

\author{
Yun Gao\footnote{
	Knight Equity Markets, L.P.,
	Jersey City, NJ 07310, USA.
	\texttt{YGao@knight.com}
	}
\and
Ioannis Kontoyiannis\footnote{
	Department of Informatics,
	Athens University of Economics and Business,
	Athens 10434, Greece.
	\texttt{yiannis@aueb.gr} 
	}
\and
Elie Bienenstock\footnote{
	Division of Applied Mathematics and
	Department of Neuroscience, Brown University, 
	Providence, RI 02912, USA.
	\texttt{elie@dam.brown.edu} 
	}
}

\begin{document}

\maketitle

\begin{abstract}
Partly motivated by entropy-estimation problems
in neuroscience, we present a detailed and extensive 
comparison between some of the most popular and
effective entropy estimation methods used in practice:
The plug-in method, four different 
estimators based on the Lempel-Ziv (LZ) family
of data compression algorithms, an estimator
based on the Context-Tree Weighting (CTW) 
method, and the renewal entropy estimator.

{\em Methodology.} Three new entropy estimators
are introduced; two new LZ-based estimators,
and the ``renewal entropy estimator,''
which is tailored to data generated by a binary 
renewal process. For two of the four LZ-based estimators,
a bootstrap procedure is described for evaluating 
their standard error, and a practical rule 
of thumb is heuristically derived for selecting 
the values of their parameters in practice.

{\em Theory.} We prove that, unlike their
earlier versions, the two new LZ-based
estimators are {\em universally} consistent, 
that is, they converge to the entropy rate 
for every finite-valued, stationary and 
ergodic process. An effective method is 
derived for the accurate approximation 
of the entropy rate of a finite-state 
HMM with known distribution. Heuristic 
calculations are presented and approximate 
formulas are derived for evaluating the bias and 
the standard error of each estimator.

{\em Simulation.} All estimators are applied
to a wide range of data generated by numerous
different processes with varying degrees of
dependence and memory. The main conclusions
drawn from these experiments include:
$(i)$~For all estimators considered, the main source 
of error is the bias. 
$(ii)$~The CTW method is repeatedly and consistently 
seen to provide the most accurate results.
$(iii)$~The performance of the LZ-based estimators 
is often comparable to that of the plug-in method.
$(iv)$~The main drawback of the plug-in method is 
its computational inefficiency; with small 
word-lengths it fails to detect longer-range structure 
in the data, and with longer word-lengths the empirical 
distribution is severely undersampled, leading to 
large biases. 
\end{abstract}

\medskip

\noindent
{\bf Keywords. } Entropy estimation, Lempel-Ziv coding, 
Context-Tree-Weighting, simulation, spike trains


\section{Introduction}

The problem of estimating the entropy of a sequence 
of discrete observations has received a lot of attention 
over the past two decades. A, necessarily incomplete,
sample of the theory that has been developed can be found in 
\cite{miller:55}\cite{basharin:59}%
\cite{grassberger}\cite{shields:1}\cite{kontoyiannis-suhov}%
\cite{treves:95}\cite{schu-grass:96}\cite{Kontoyiannis:96}\cite{kasw}%
\cite{darbellay:99}\cite{victor:00}\cite{antos-K:01}\cite{paninski:03}%
\cite{C-K-Verdu:04} 
and the references therein. Examples of numerous 
different applications are contained in the above list,
as well as in
\cite{brown:IBM}\cite{chen-reif:1}\cite{adi-dna}%
\cite{Stevens:96}\cite{teahan:cleary}%
\cite{Strong:98}\cite{Suzuki:99}\cite{LW:99}%
\cite{levene-loizou:00}\cite{Pamela:00}\cite{london}%
\cite{Bhumbra:04}\cite{nemenman:04}.

Information-theoretic methods 
have been particularly widely used in neuroscience, 
in a broad effort to analyze and understand the fundamental 
information-processing tasks performed by 
the brain. In many of these these studies, 
the entropy is adopted as the main measure 
for describing the amount of information 
transmitted between neurons. There, 
one of the most basic tasks is to identify 
appropriate methods for
quantifying this information, in other words, 
to estimate the entropy of spike trains 
recorded from live animals; see, e.g.,
\cite{Stevens:96}\cite{warland:97}\cite{Strong:98}\cite{Pamela:00}%
\cite{london}\cite{kennel}\cite{paninski:03}\cite{Bhumbra:04}%
\cite{nemenman:04}\cite{shlens:07}.

Motivated, in part, by the application
of entropy estimation techniques to such
neuronal data, we present
a systematic and extensive comparison,
both in theory and via simulation,
between several of the most commonly used 
entropy estimation techniques. This work serves,
partly, as a more theoretical companion 
to the experimental work and results presented 
in \cite{neuro-nips}\cite{gao:thesis}\cite{neuro-isit}.
There, entropy estimators were applied to the 
spike trains of 28 neurons recorded simultaneously 
for a one-hour period from the primary motor and dorsal 
premotor cortices (MI, PMd) of a monkey. The purpose of those
experiments was to examine the effectiveness 
of the entropy as a statistic, and its utility
in revealing some of the underlying 
structural and statistical characteristics
of the spike trains. In contrast, our main
aim here is to examine the performance of
several of the most effective entropy estimators,
and to establish fundamental properties for their
applicability, such as rigorous estimates for their
convergence rates, bias and variance.
In particular, since (discretized) spike trains
are typically represented as binary sequences \cite{spikes:book},
some of our theoretical results and all 
of our simulation experiments are focused
on binary data.

Section~2 begins with a description of the
entropy estimators we consider. The simplest
one is the {\em plug-in} or {\em maximum likelihood}
estimator, which consists of first calculating 
the empirical frequencies of all words of a fixed length 
in the data, and then computing the entropy 
of this empirical distribution.
For obvious computational reasons, the plug-in is
ineffective for word-lengths beyond 10 or 20, and hence 
it cannot take into account any potential longer-time 
dependencies in the data. 

A popular approach for overcoming this 
drawback is to consider entropy estimators 
based on ``universal'' data compression algorithms, 
that is, algorithms which are known to achieve 
a compression ratio equal to the entropy, for data 
generated by processes which may possess arbitrarily 
long memory, and without any prior knowledge about 
the distribution of the underlying process.
Since, when trying to estimate the entropy,
the actual compression task is irrelevant,
many entropy estimators have been developed 
as modifications of practical compression
schemes. Section~\ref{subsec:lz} describes 
two well-known such entropy estimators \cite{kasw}, 
which are based on the Lempel-Ziv (LZ) family of 
data compression algorithms \cite{ziv-lempel:1}\cite{ziv-lempel:2}.
These estimators are known to be {\em consistent}
(i.e., to converge to the correct value for the entropy) 
only under certain restrictive conditions on 
the data. We introduce two {\em new} LZ-based 
estimators, and we prove in Theorem~\ref{thm1} that, 
unlike the estimators in \cite{kasw}, they are
consistent under essentially minimal conditions,
that is, for data generated by any stationary
and ergodic process.

Section~\ref{subsec:biasvarlz} contains an analysis,
partly rigorous and partly in terms of heuristic computations,
of the rate at which the bias and the variance of 
each of the four LZ-based estimators converges
to zero. A bootstrap procedure is developed for 
empirically estimating the standard error of two 
of the four LZ-based estimators, and a practical 
rule-of-thumb is derived for selecting the values 
of the parameters of these estimators in practice.

Next, in Section~\ref{subsec:ctwtheory}
we consider an entropy estimator based
on the Context-Tree Weighting (CTW) algorithm
\cite{willems-shtarkov-tjalkens:95}\cite{willems:96}\cite{willems:98}.
[In the neuroscience literature, a similar proceedure
has been applied in \cite{kennel} and \cite{london}.]
The CTW, also originally developed for data compression,
can be interpreted as a Bayesian estimation procedure.
After a brief description, we explain that it
is consistent for data generated by any stationary
and ergodic process and show that
its bias and variance are, in a sense, as small 
as can be.\footnote{The 
	CTW also has another feature which, although important
	for applied statistical analyses such as those
	reported in connection with our experimental 
	results in \cite{gao:thesis}\cite{neuro-isit},
	will not be explored in the present work:
	A simple modification of the algorithm can be
	used to compute the maximum a posteriori probability
	tree model for the data; see Section~\ref{subsec:ctwtheory} 
	for some details.}

Section~\ref{sec:simu} contains the results
of an extensive simulation study, where the 
various entropy estimators are applied to data
generated from numerous different types of
processes, with varying degrees of dependence
and memory. In Section~\ref{subsec:simutrueentropy},
after giving brief descriptions of all these data models,
we present (Proposition~\ref{prop1}) a method for 
accurately approximating 
the entropy rate of a Hidden Markov Model (HMM);
recall that HMMs are not known to admit
closed-form expressions for their entropy rate.
Also, again partly motivated by neuroscience 
applications, we introduce another new entropy
estimator, the {\em renewal entropy estimator}, 
which is tailored to binary data generated by 
renewal processes.

Section~\ref{subsec:conver} contains a detailed
examination of the bias and variance of the four
LZ-based estimators and the CTW algorithm. There,
our earlier theoretical predictions are largely
confirmed. Moreover, it is found that two of the
four LZ-based estimators are consistently more
accurate than the other two.

Finally, Section~\ref{subsec:comp} contains
a systematic comparison of the performance of all 
of the above estimators on different types of 
simulated data. Incidental comparisons between
some of these methods on limited data sets have
appeared in various places in the literature,
and questions have often been raised regarding 
their relative merits. One of the {\em main goals}
of the work we report here is to offer clear 
resolutions for many of these issues.
Specifically, in addition to the points
mentioned up to now, some of the main conclusion 
that we draw from the simulation results 
of Section~\ref{sec:simu} (these and more 
are collected in Section~\ref{sec:concl})
can be summarized as follows:
\begin{itemize}
\item
Due its computational inefficiency,
the plug-in estimator is 
the least reliable method, in contrast to
the LZ-based estimators and the CTW, 
which naturally incorporate dependencies 
in the data at much larger time scales.
\item
The most effective estimator is the
CTW method. Moreover, for the CTW as
well as for all other estimators, 
the main source of error is the bias
and not the variance.
\item
Among the four LZ-based estimators, 
the two most efficient ones are those 
with increasing window sizes, $\hat{H}_n$ of
\cite{kasw} and $\tilde{H_n}$ introduced
in Section~\ref{subsec:lz}.
Somewhat surprisingly, in several of the
simulations we conducted the performance 
of the LZ-based estimators appears to be
very similar to that of the plug-in method.
\end{itemize}


\section{Entropy Estimators and Their Properties}
\label{sec:methods}

This section contains a detailed description of the
entropy estimators that will be applied to simulated data
in Section~\ref{sec:simu}. After some basic definitions
and notation in Section~\ref{subsec:H}, the following
four subsections contain the definitions of the estimators
together with a discussion of their statistical properties,
including conditions for consistency, and estimates of their
bias and variance.

\subsection{Entropy and Entropy Rate}
\label{subsec:H}
Let $X$ be a random variable or
random vector, taking values in an arbitrary
finite set $A$, its {\em alphabet}, and with 
distribution $p(x)=\Pr\{X=x\}$ for $x\in A$.
The {\em entropy of $X$} \cite{cover:book} is defined as,
\[
H(X)=H(p)=-\sum_{x\in A} p(x)\log p(x),
\]
where, throughout the paper, $\log$ denotes the logarithm 
to base 2, $\log_2$.
A random process $\Xp=\{\ldots,X_{-1},X_0,X_1,X_2,\ldots\}$ 
with alphabet $A$ is a sequence of random variables $\{X_n\}$
with values in $A$. We write $X_i^j=(X_i,X_{i+1},\ldots,X_j)$
for a (possibly infinite) contiguous segment of the process,
with $-\infty\leq i\leq j\leq \infty$,
and $x_i^j=(x_i,x_{i+1},\ldots,x_j)$ for a specific
realization of $X_i^j$, so that $x_i^j$ is an element 
of $A^{j-i+1}$.
The {\em entropy rate} $H=H(\Xp)$, or ``per-symbol'' entropy,
of $\Xp$ is the asymptotic rate at which the entropy of $X_1^n
$ changes with $n$, 
\be
H=H(\Xp)=\lim_{n \rightarrow \infty} \frac{1}{n} H(X_1,X_2,\ldots,X_n),
\label{eq:ERdefn}
\ee
whenever the limit exists, where $H(X_1,X_2,\ldots,X_n)$ 
is the entropy of the jointly distributed random variables 
$X_1^n=(X_1,X_2,\ldots,X_n)$. Recall \cite{cover:book} that 
for a {\em stationary} process (i.e., a process such 
that the distribution of every finite block $X_{n+1}^{n+k}$ 
of size $k$ has the same distribution, say $p_k$, independently 
of its position $n$), the entropy rate exists and equals,
$$H=H(\Xp) = \lim_{n \rightarrow \infty} H(X_n \,|\, X_{n-1},\ldots,X_2,X_1),$$
where the {\em conditional entropy}
$H(X_n\,|\,X_1^{n-1})=H(X_n \,|\, X_{n-1},\ldots,X_2,X_1)$ 
is defined as,
\ben
H(X_n\,|\,X_1^{n-1})
&=&-\sum_{x_1^n\in A^n} p_n(x_1^n)\log 
	\Pr\{X_n=x_n\,|\,X_1^{n-1}=x_1^{n-1}\}\\
&=&-\sum_{x_1^n\in A^n} p_n(x_1^n)\log 
	\frac{p_n(x_1^n)}{p_{n-1}(x_1^{n-1})}.
\een

As mentioned in the introduction, much of the rest
of the paper will be devoted to binary data 
produced by processes $\Xp$ with alphabet $A=\{0,1\}$.

\subsection{The Plug-in Estimator}
Perhaps the simplest and most straightforward
estimator for the entropy rate 
is the so-called plug-in 
estimator. Given a data
sequence $x_1^n$ of length $n$, and an arbitrary
string, or ``word,'' $y_1^w\in A^w$ of length $w<n$,
let $\hat{p}_w(y_1^w)$ denote the empirical probability 
of the word $y_1^w$ in $x_1^n$; that is,
$\hat{p}_w(y_1^w)$ is the frequency with which 
$y_1^w$ appears in $x_1^n$. If the data are produced
from a stationary and ergodic\footnote{Recall 
	that ergodicity is simply the assumption
	that the law of large numbers holds in its 
	general form (the ergodic theorem);
	see, e.g., \cite{shields:book} for details.
	This assumption is natural and, in 
	a sense, minimal, in that it is hard to
	even imagine how any kind of statistical
	inference would be possible if we cannot
	even rely on taking long-term averages.
	}
process, 
then the law of large numbers
guarantees that, for fixed $w$ and large $n$,
the empirical distribution  $\hat{p}_w$ will
be close to the true distribution $p_w$,
and therefore a natural estimator for the entropy
rate based on (\ref{eq:ERdefn}) is:
\[
\hat{H}_{n,w,{\rm plug-in}} 
= \frac{1}{w}H(\hat{p}_w)
= -\frac{1}{w} \sum_{y_1^w\in A^w} \hat{p}_w(y_1^w) \log\hat{p}_w(y_1^w).
\]
This is the {\em plug-in estimator with word-length $w$}. 
Since the empirical 
distribution is also the maximum likelihood estimate 
of the true distribution, this is also often 
referred to as the
{\em maximum-likelihood} entropy estimator.

Suppose the process $\Xp$ is stationary and ergodic.
Then, taking $w$ large enough for 
$\frac{1}{w}H(X_1^w)$ to be acceptably close to $H$
in (\ref{eq:ERdefn}), and assuming the number of samples
$n$ is much larger than $w$ so that the empirical 
distribution of order $w$ is close to the true distribution,
the plug-in estimator $\hat{H}_{n,w,{\rm plug-in}}$ 
will produce an accurate 
estimate for the entropy rate. But, among other difficulties,
in practice this leads to enormous computational 
problems because the number of all possible words 
of length $w$ grows {\em exponentially} with $w$. 
For example, even for the
simple case of binary data with a modest word-length of 
$w=30$, the number of possible strings $y_1^w$
is $2^{30}$, which in practice means that the
estimator would either require astronomical 
amounts of data to estimate $\hat{p}_w$ accurately,
or it would be severely undersampled. See also 
\cite{paninski:04} and the references therein
for a discussion of the undersampling problem.

Another drawback of the plug-in estimator 
is that it is hard to quantify its bias and to correct
for it. For any fixed word-length $w$, it is easy to 
see that the bias
$E\big[\hat{H}_{n,w,{\rm plug-in}}\big]-\frac{1}{w}H(X_1^w)$
is always negative \cite{antos-K:01}\cite{paninski:03},
whereas the difference between the
$w$th-order per-symbol entropy and the entropy rate,
$\frac{1}{w}H(X_1^w)-H(\Xp),$ is always 
nonnegative. Still, there have been
numerous extensive studies on calculating this bias
and on developing ways to correct for it;
see \cite{Strong:98}\cite{paninski:03}\cite{warland:97}%
\cite{Pamela:00}\cite{Stevens:96}\cite{nemenman:04} 
and the references therein.

\subsection{The Lempel-Ziv Estimators}
\label{subsec:lz}

An intuitively appealing and popular way of estimating 
the entropy of discrete data with possibly long memory,
is based on the use 
of so-called {\em universal} data compression algorithms. 
These are algorithms that are known to be able to 
optimally compress data from an arbitrary process
(assuming some broad conditions are satisfied), where
optimality means that the compression ratio they achieve
is asymptotically equal to the entropy rate of the 
underlying process -- although the statistics of this 
process are {\em not} assumed to be known {\em a priori}.
Perhaps the most commonly used methods in this context 
are based on a family of compression schemes known as 
Lempel-Ziv (LZ) algorithms; see, e.g., 
\cite{ziv-lempel:1}\cite{ziv-lempel:2}\cite{wyner-ziv:1}.

Since the entropy estimation task is simpler than that
of actually compressing the data, several modified versions
of the original compression algorithms have been proposed
and used extensively in practice. All these methods are
based on the calculation of the lengths of certain repeating 
patterns in the data. Specifically, given a data realization 
$\xp=(\ldots,x_{-1},x_0,x_1,x_2,\ldots)$,
for every position $i$ in $\xp$ and any ``window length'' $n\geq 1$,
consider the length $\ell$ of the longest
segment $x_i^{i+\ell-1}$ in the data starting at $i$ which also appears 
in the window $x_{i-n}^{i-1}$ of length $n$ preceding position $i$. 
Formally, define $L_i^n=L_i^n(\xp)=
L_i^n(x_{i-n}^{i+n-1})$ 
as $1+$ [that longest match-length]:
\begin{eqnarray*}
L_i^n 
&=& L_i^n(\xp) 
\;=\; 
	L_i^n(x_{i-n}^{i+n-1})\\
&=&
	1+ 
	\max\{0\leq \ell\leq n: 
	x_i^{i+\ell-1}
	=
	x_j^{j+\ell-1}
	\mbox{ for some } i-n \le j \le i-1 \}.
\end{eqnarray*}

Suppose that the process $\Xp$ 
is stationary and ergodic,
and consider the random match-lengths 
$L_i^n=L_i^n(X_{i-n}^{i+n-1})$.
In \cite{wyner-ziv:1}\cite{ornstein-weiss:2}
it was shown that,
for any fixed position $i$,
the match-lengths grow logarithmically
with the window size $n$, and in fact,
\begin{eqnarray}
\frac{L_i^n}{\log n} \to \frac{1}{H}\;\;\;\mbox{as}\;\;\;
n\to\infty,\;\;\;\mbox{with probability 1},
\label{eq:OW}
\end{eqnarray}
where $H$ is the entropy rate of the process.
This result suggests that the quantity
$(\log n)/L_i^n$ can be used as an entropy
estimator, and, clearly, in order to make
more efficient use of the data and 
reduce the variance, it would be more
reasonable to look at the average
value of various match-lengths $L_i^n$
taken at different positions $i$;
see the discussion
in \cite{kasw}.
To that effect, the following two estimators
are considered in \cite{kasw}.
Given a data realization $\xp=x_{-\infty}^\infty$,
a window length $n\geq 1$, and a number of matches $k\geq 1$,
the {\em sliding-window LZ estimator}
$\hat{H}_{n,k} = 
\hat{H}_{n,k}(\xp) = 
\hat{H}_{n,k}(x_{-n+1}^{n+k-1})$
is defined by,
\begin{equation}
\hat{H}_{n,k} = 
\left[ \frac{1}{k}\sum_{i=1}^k \frac{L_i^n}{\log n} \right]^{-1}.
\label{eq:Hhat}
\end{equation}
Similarly, the {\em increasing-window LZ estimator}
$\hat{H}_{n} = 
\hat{H}_{n}(\xp) = 
\hat{H}_{n}(x_0^{2n-1})$
is defined by,
\begin{equation}
\hat{H}_{n} = 
\left[ \frac{1}{n}\sum_{i=2}^n \frac{L_i^i}{\log i} \right]^{-1}.
\label{eq:Hhat_whole}
\end{equation}

The difference between the two estimators in 
(\ref{eq:Hhat}) and (\ref{eq:Hhat_whole})
is that 
$\hat{H}_{n,k}$ 
uses a fixed window length, 
while  
$\hat{H}_{n}$ 
uses the entire history 
as its window, so that the window length increases  
as the matching position moves forward.

In \cite{kasw} it is shown that, 
under appropriate conditions, 
both estimators $\hat{H}_{n,k}$ and $\hat{H}_{n}$ 
are consistent, in that they converge to the entropy 
rate of the underlying process with probability 1,
as $n,k\to\infty$.
Specifically, it is assumed that the
process is stationary and ergodic, that it takes
on only finitely many values, and that
it satisfies the {\em Doeblin Condition} (DC).
This condition says that there is a
finite number of steps, say $r$, in the
process, such that, after $r$ time steps,
no matter what has occurred before, anything
can happen with positive probability:

\begin{quote}
{\bf Doeblin Condition (DC). }
There exists an integer 
$r \ge 1$ and a real number $\beta>0$ 
such that,
$$\mbox{Pr}(X_r=a\,|\,X_{-\infty}^0)>\beta,$$
for all $a\in A$ and with probability one in the 
conditioning, i.e., for almost all 
semi-infinite realizations of the past 
$X_{-\infty}^0=(\,X_0,X_{-1},\ldots)$.
\end{quote}

Condition (DC) has the advantage that it is
not quantitative -- the values of $r$ and $\beta$
can be arbitrary -- and, therefore, for specific
applications it is fairly easy to see whether it
is satisfied or not.
But it is restrictive, and, as
it turns out, it can be avoided altogether 
if we consider a modified version of 
the above two estimators.

To that end, we define two new estimators
$\tilde{H}_{n,k}$ and $\tilde{H}_{n}$ as follows.
Given $\xp=x_{-\infty}^\infty$, $n$ and $k$ as above,
define the new sliding-window estimator
$\tilde{H}_{n,k} =
\tilde{H}_{n,k}(\xp) =
\tilde{H}_{n,k}(x_{-n+1}^{n+k-1})$,
\begin{equation}
\tilde{H}_{n,k} = \frac{1}{k}\sum_{i=1}^k \frac{\log n}{L_i^n},
\label{eq:Htilde}
\end{equation}
and the new increasing-window estimator
$\tilde{H}_{n} =
\tilde{H}_{n}(\xp) =
\tilde{H}_{n}(x_0^{2n-1})$ as,
\begin{equation}
\tilde{H}_{n} = \frac{1}{n}\sum_{i=2}^n \frac{\log i}{L_i^i}.
\label{eq:Htilde_whole}
\end{equation}

Below some basic properties of 
these four estimators are established, 
and conditions are given for their
asymptotic consistency.
Parts~(i) and~(iii) of Theorem~\ref{thm1} are new;
most of part~(ii) is contained in \cite{kasw}.

\begin{theorem} {\sc [Consistency of LZ-type Estimators]}
\label{thm1}
\begin{itemize}
\item[{\em (i)}]
	When applied to an {\em arbitrary} data string,
	the estimators defined in 
	{\em (\ref{eq:Hhat})--(\ref{eq:Htilde_whole})} 
	always satisfy,
	$$\hat{H}_{n,k} \leq \tilde{H}_{n,k}
	\;\;\;\mbox{and}\;\;\;
        \hat{H}_{n} \leq \tilde{H}_{n},$$
	for any $n,k$.
\item[{\em (ii)}]
	The estimators 
        $\hat{H}_{n,k}$ and $\hat{H}_{n}$
	are consistent when applied to data
	generated by a finite-valued, 
	stationary, ergodic process that
        satisfies Doeblin's condition {\em (DC)}.
	With probability one we have:
        $$\hat{H}_{n,k} \to H,
	\;\;\hat{H}_{n} \to H,\;\;\;\mbox{as }k,n\to\infty.$$
\item[{\em (iii)}]
	The estimators 
        $\tilde{H}_{n,k}$ and $\tilde{H}_{n}$
	are consistent when applied to data
	generated by an {\em arbitrary finite-valued, 
	stationary, ergodic process},
	even if {\em (DC)} does not hold.
	With probability one we have:
        $$\tilde{H}_{n,k} \to H,\;\;\tilde{H}_{n} 
	\to H,\;\;\;\mbox{as }k,n\to\infty.$$
\end{itemize}
\end{theorem}

Note that parts~(ii) and (iii) do not specify
the manner in which $n$ and $k$ go to infinity. 
The results are actually valid in the following cases:
\begin{enumerate}
\item
If the two limits as $n$ and $k$ tend to infinity are taken 
separately, i.e., first $k\to\infty$ and then
$n\to\infty$, or vice versa;
\item
If $k\to\infty$ and $n=n_k$ varies with $k$ 
in such a way that $n_k\to\infty$ 
as $k\to\infty$;  
\item
If $n\to\infty$ and $k=k_n$ varies with $n$
in such a way that it increases to infinity
as $n\to\infty$;
\item
If $k$ and $n$ both vary arbitrarily in such a way 
that $k$ stays bounded and $n\to\infty$.
\end{enumerate}

\vspace{0.1in}

\noindent
{\bf Proof. } 
{\sc Part} (i). An application 
of Jensen's inequality to the convex function
$x\mapsto 1/x$,
with respect to the uniform distribution
$(1/k,\ldots,1/k)$ on the set $\{1,2,\ldots,k\}$,
yields,
\begin{eqnarray*}
\tilde{H}_{n,k}
\;=\;
	\sum_{i=1}^k\frac{1}{k}\frac{\log n}{L_i^n}
	\;=\;
	\sum_{i=1}^k\frac{1}{k}\Big[\frac{L_i^n}{\log n}\Big]^{-1}
	\;\geq\;
	\Big[\sum_{i=1}^k\frac{1}{k}\frac{L_i^n}{\log n}\Big]^{-1}
	\;=\;\hat{H}_{n,k},
\end{eqnarray*}
as required. The proof of the second assertion is similar.

\vspace{0.1in}

\noindent
{\sc Part} (ii). The results here are, for the most part, 
proved in \cite{kasw}, where it is 
established that $\hat{H}_n\to H$ 
and $\hat{H}_{n,n}\to H$ as $n\to\infty$,
with probability one.
So it remains to show that 
$\hat{H}_{n,k}\to H$ as $n,k\to\infty$
in each of the four cases stated above.

For case~1 observe that, with probability 1,
\be
\lim_k\lim_n \hat{H}_{n,k}
\;=\;
	\lim_k\left[\frac{1}{k}\sum_{i=1}^k
	\lim_n \frac{L_i^n}{\log n}\right]^{-1}
\;\eqa\;
	\lim_k\left[\frac{1}{k}\sum_{i=1}^k\frac{1}{H}\right]^{-1}
\;=\; H,
\label{eq:double}
\ee
where $(a)$ follows from (\ref{eq:OW}). To reverse the limits,
we define, for each fixed $n$, a new process 
$\{Y_i^{(n)}\}= 
\{\ldots,Y_{-1}^{(n)},Y_{0}^{(n)},Y_{1}^{(n)},Y_{2}^{(n)},\ldots\}$
by letting $Y^{(n)}_i=L_i^n/(\log n)$ for each $i$. Then
the process $\{Y^{(n)}_i\}$ is itself stationary and ergodic.
Recalling also from \cite{kasw} that the convergence
in (\ref{eq:OW}) takes place not only with probability one but
also in $L^1$, we may apply the ergodic theorem to obtain
that, with probability 1,
\ben
\lim_n\lim_k \hat{H}_{n,k}
&=&
	\lim_n\left[\lim_k\frac{1}{k}\sum_{i=1}^k
	Y^{(n)}_i\right]^{-1}\\
&\eqb&
	\lim_n\;[E(Y^{(n)}_1)]^{-1}
\;=\;
	\left[\lim_nE\Big(\frac{L_1^n}{\log n}\Big)\right]^{-1}\\
&\eqc&
	H,
\een
where $(b)$ follows by the ergodic theorem and $(c)$ from the
$L^1$ version of (\ref{eq:OW}).

The proof of case~2 is identical to the case $k=n$
considered in \cite{kasw}. In case~3,
since the sequence $\{k_n\}$ is increasing,
the limit of $H_{k_n,n}$ reduces to a subsequence 
of the corresponding limit in case~2 
upon considering the inverse sequence $\{n_k\}$.

Finally for case~4, recall from (\ref{eq:double}) that,
$$\lim_{n} H_{k,n}\;=\;H\;\;\;\mbox{with prob.\ 1},$$
for any fixed $k$. Therefore the same will
hold with a varying $k$, as long as it
varies among finitely many values.

\vspace{0.1in}

\noindent
{\sc Part} (iii). The proofs of the consistency
results for $\tilde{H}_n$ and $\tilde{H}_{n,k}$
can be carried out along the same lines as
the proofs of the corresponding results in \cite{kasw},
together with their extensions as in Part~(ii) above.
The only difference is in the main technical step,
namely, the verification of a uniform integrability
condition. In the present case, what is needed is 
to show that,
\be
E\Big\{\sup_{n\geq 1}\frac{\log n}{L_1^n}\Big\}<\infty.
\label{eq:L1dom}
\ee
This is done in the following lemma.
\hfill $\Box$

\begin{lemma}
Under the assumptions of part~{\em (iii)} of the theorem, 
the $L^1$-domination condition {\em (\ref{eq:L1dom})} 
holds true.
\end{lemma}

\noindent
{\bf Proof. }
Given a data realization 
$\xp=(\ldots,x_{-2},x_{-1},x_0,x_1,x_2,\ldots)$
and an $m\geq 1$, the recurrence time $R_m$ is
defined as the first time the substring $x_1^m$ appears
again in the past. More precisely, $R_m$ is the number of
steps to the left of $x_1^m$ we have to look in order
to find a copy of $x_1^m$:
\begin{eqnarray*}
R_m=R_m(\xp)\;=\;
R_m(x_{-\infty}^m)
\;=\; \inf
	\{k\geq 1: 
	x_1^m
	=
	x_{-k+1}^{-k+m}
	\}.
\end{eqnarray*}
For any such realization $\xp$ and any $n\geq 1$, 
if we take $m=L_1^n$,
then by the definitions of $R_m$ and $L_1^n$ it follows that,
$R_m>n,$ which implies,
$$\frac{\log R_m}{m}>\frac{\log n}{L_1^n},$$
and thus it is always the case that,
$$ \sup_n\frac{\log n}{L_1^n}<\sup_m \frac{\log R_m}{m}.$$
Therefore, to establish (\ref{eq:L1dom}) it suffices
to prove:
\be
E\Big\{ \sup_m\frac{\log R_m}{m} \Big\}<\infty.
\label{eq:L1dom2}
\ee
To that end, we expand this expectation as,
\ben
E\Big\{ \sup_m\frac{\log R_m}{m} \Big\}
&\leq&
	\sum_{k\geq 0}\Pr\Big\{
	\sup_m\frac{\log R_m}{m} \geq k\Big\}\\
&\leq&
	K +
	\sum_{k\geq K}\Pr\Big\{
	\sup_m\frac{\log R_m}{m} \geq k\Big\}\\
&\leq&
	K +
	\sum_{k\geq K}\sum_{m\geq 1}
	\Pr\Big\{
	\frac{\log R_m}{m} \geq k\Big\},
\een
where $K$ is an arbitrary integer to be chosen later.
Applying Markov's inequality,
\be
E\Big\{ \sup_m\frac{\log R_m}{m} \Big\}
&\leq&
	K +
	\sum_{k\geq K}
	\sum_{m\geq 1}
	E(R_m) 2^{-mk}.
\label{eq:markov}
\ee
To calculate the expectation of $R_m$, suppose that
the process $\Xp$ takes on $\alpha=|A|$ possible values, so
that there are $\alpha^m$ possible strings $x_1^m$ of
length $m$.
Now recall Kac's theorem 
\cite{wyner-ziv:1}
which states that
$E(R_m\,|\,X_1^m=x_1^m)=1/\Pr\{X_1^m=x_1^m\}$,
from which it follows that,
\be
E(R_m)=\sum_{x_1^m}E(R_m\,|\,X_1^m=x_1^m)\cdot\Pr\{X_1^m=x_1^m\} = \alpha^m.
\label{eq:kac}
\ee
Combining (\ref{eq:markov}) and (\ref{eq:kac}) yields,
\ben
E\Big\{ \sup_m \frac{\log R_m}{m} \Big\}
&\leq&
	K +
	\sum_{k\geq K}
	\sum_{m\geq 1}
	2^{-m(k-\log \alpha)}\\
&=&
	K +
	\sum_{k\geq K}
	\frac{2^{-(k-\log \alpha)}}
	{1-2^{-(k-\log \alpha)}}\\
&=&
	K +
	\sum_{k\geq K}
	\frac{1}
	{\frac{2^k}{\alpha}-1}
\;<\;\infty,
\een
where we choose $K>\log \alpha$.
This establishes (\ref{eq:L1dom2})
and completes the proof.
\hfill $\Box$

\subsection{Bias and Variance of the LZ-based Estimators}
\label{subsec:biasvarlz}

In practice, when applying the sliding-window 
LZ estimators $\hat{H}_{n,k}$ or $\tilde{H}_{n,k}$
on finite data strings, 
the values of the parameters $k$ and $n$ 
need to be chosen, so that $k+n$ is
approximately equal to the total data
length. This presents the following
dilemma: Using a long window size $n$, the
estimators are more likely to capture the 
longer-term trends in the data, but, as shown
in \cite{pittel:1}\cite{szpankowski:1},
the match-lengths $L_i^n$ starting
at different positions $i$ have large
fluctuations. So a large window size
$n$ and a small number of matching
positions $k$ will give estimates
with high variance. On the other hand,
if a relatively small value for $n$ is chosen
and the estimate is an average over a
large number of positions $k$,
then the variance will be reduced at
the cost of increasing the bias,
since the expected value of $L_i^n/\log n$
is known to converge to $1/H$ very slowly 
\cite{wyner-wyner}.

Therefore $n$ and $k$ need to be chosen in a way
such that the above bias/variance trade-off is balanced.
From the earlier theoretical results of
\cite{pittel:1}\cite{szpankowski:1}%
\cite{szpankowski:93b}\cite{wyner-wyner}\cite{wyner-ziv-wyner}
it follows that, under appropriate conditions, 
the bias is approximately of the order $O(1/\log n)$, 
whereas from the central limit theorem it is easily 
seen that the variance is approximately of order
$O(1/k)$. This indicates that the relative values
of $n$ and $k$ should probably be chosen 
to satisfy $k\approx O((\log n)^2).$

Although this is a useful general guideline,
we also consider the problem of empirically 
evaluating the relative estimation error 
on particular data sets. Next we outline 
a bootstrap procedure, which
gives empirical estimates of the variance
$\hat{H}_{n,k}$ and $\tilde{H}_{n,k}$;
an analogous method was used for the estimator
$\hat{H}_{n,k}$ in \cite{Suzuki:99}, in the
context of estimating the entropy of whale songs.

Let $L$ denote
the sequence of match-lengths 
$L=(L_1^n,,L_2^n,\ldots,L_k^n)$
computed directly from the data, as in the definitions
of $\hat{H}_{n,k}$ and $\tilde{H}_{n,k}$.
Roughly speaking, the proposed procedure is 
carried out in three steps:
First, we sample with replacement from $L$
in order to obtain many pseudo-time series with the same 
length as $L$; then we compute new entropy estimates 
from each of the new sequences using 
$\hat{H}_{n,k}$ or $\tilde{H}_{n,k}$;
and finally we estimate the variance of the initial 
entropy estimates as the sample variance of the new
estimates.  The most important step is the sampling,
since the elements of sequence $(L_1^n,\ldots,L_k^n)$
are {\em not} independent. In order to 
maintain the right form of dependence,
we adopt a version of the {\em stationary bootstrap} 
procedure of \cite{Politis:94}. The basic idea is, 
instead of sampling individual $L_i^n$'s from
$L$, to sample whole blocks with random lengths.
The choice of the distribution of their lengths 
is made in such a way as to guarantee that they
are typically
long enough to maintain sufficient dependence
as in the original sequence. The results in 
\cite{Politis:94} provide conditions
which justify the application 
of this procedure.

The details of the three steps above are as follows:
First, a random position $j\in\{1,2,\ldots,k\}$ 
is selected uniformly at random,
and a random length $T$ is chosen with 
geometric distribution with mean $1/p$ 
(the choice of $p$ is discussed below). Then the
block of match-lengths $(L_j^n,L_{j+1}^n,\ldots,L_{j+T-1})$
is copied from $L$, and the same process is repeated
until the concatenation $L^*$ of the sampled blocks 
has length $k$. This gives the first bootstrap sample. 
Then the whole process is repeated to generate 
a total of $B$ such blocks $L^{*1},L^{*2},\ldots,L^{*B}$,
each of length $k$. From these we calculate new entropy 
estimates $\hat{H}^\ast(m)$ 
or $\tilde{H}^\ast(m),$ 
for $m=1,2,\ldots,B$, according to the 
definition of the entropy estimator being used,
as in (\ref{eq:Hhat}) or (\ref{eq:Htilde}),
respectively; the choice of the number $B$ of blocks
is discussed below.
The bootstrap estimate of the variance of $\hat{H}$ 
is,
$$\hat{\sigma}^2 = \frac{1}{B-1}
\sum_{m=1}^B [\hat{H}^\ast(m)-\hat{\mu}]^2,$$
where $\hat{\mu} = B^{-1}\sum_{m=1}^B \hat{H}^\ast(m)$;
similarly for $\tilde{H}$. 

The choice of the parameter $p$ depends on the 
length of the memory of the match-length sequence 
$L=(L_1^n,L_2^n,\ldots,L_k^n)$;
the longer the memory, the larger 
the blocks need to be, therefore, the smaller
the parameter $p$. In practice,
$p$ is chosen by studying the autocorrelogram of $L$,
which is typically decreasing with the lag:
We choose an appropriate cutoff threshold, 
take the corresponding lag to be the average 
block size, and choose $p$ as the
reciprocal of that lag. Finally, the number
of blocks $B$ is customarily chosen
large enough so that the histogram of the 
bootstrap samples $\hat{H}^\ast(1),\hat{H}^\ast(2),
\ldots,\hat{H}^\ast(B)$ ``looks'' approximately
Gaussian. Typical values used in applications
are between $B=500$ and $B=1000$. In all our 
experiments in \cite{gao:thesis}\cite{neuro-isit}
and in the results 
presented in the following section
we set $B=1000$, which, as discussed below,
appears to have been sufficiently large for 
the central limit theorem to apply 
to within a close approximation.

\subsection{Context-Tree Weighting}
\label{subsec:ctwtheory}
One of the fundamental ways in which the entropy rate
arises as a natural quantity, is in the Shannon-McMillan-Breiman
theorem \cite{cover:book}; it
states that, for any stationary and ergodic
process $\Xp=\{\ldots,X_{-1},X_0,X_1,X_2,\ldots\}$ 
with entropy rate $H$,
\be
-\frac{1}{n}\log p_n(X_1^n)\to H,
\;\;\;\;\mbox{with prob.\ 1, as}\;n\to\infty,
\label{eq:SMBT}
\ee
where $p_n(X_1^n)$ denotes the (random) probability of
the random string $X_1^n$. This suggests that,
one way to estimate $H$
from a long realization $x_1^n$ of $\Xp$,
is to first somehow estimate its probability
and then use the estimated probability $\hat{p}_n(x_1^n)$
to obtain an estimate for the entropy rate via,
\be
\hat{H}_{n,{\rm est}}=-\frac{1}{n}\log \hat{p}_n(x_1^n).
\label{eq:SMBTest}
\ee

The Context-Tree Weighting (CTW) algorithm
\cite{willems-shtarkov-tjalkens:95}\cite{willems:96}\cite{willems:98}
is a method, originally developed in the context of
data compression, which can be interpreted 
as an implementation of hierarchical Bayesian
procedure for estimating the probability
of a string generated by a binary ``tree process.''\footnote{The
	CTW algorithm is a general method with various 
	extensions, which go well beyond the basic version 
	described here. Some of these extensions are mentioned
	later in this section.}
The details of the precise way in which 
the CTW operates can be found in 
\cite{willems-shtarkov-tjalkens:95}\cite{willems:96}\cite{willems:98};
here we simply give a brief overview of what 
(and not how) the CTW actually computes.
In order to do that, we first need to describe
tree processes.

A {\em binary tree process of depth $D$} is a binary process
$\Xp$ with a distribution defined in terms of a 
{\em suffix set} $S$, consisting of binary strings of length no longer
than $D$, and a parameter vector $\thetap=(\theta_s\;;\;s\in S)$,
where each $\theta_s\in[0,1]$.
The suffix set $S$ is assumed to be complete and proper,
which means that any semi-infinite binary string
$x_{-\infty}^0$ has exactly one suffix $s$ is $S$, i.e.,
there exists exactly one $s\in S$ such that 
$x_{-\infty}^0$ can be written as $x_{-\infty}^{-k} s$,
for some integer $k$.\footnote{
	The name {\em tree process} comes 
	from the fact that the suffix set $S$ can be 
	represented as a binary {\em tree}.
	}
We write $s=s(x_{-\infty}^0)\in S$ for this unique
suffix.

Then the distribution of $\Xp$ is specified by 
defining the conditional probabilities,
$$\Pr\{X_{n+1}=1\,|\,X_{-\infty}^n=x_{-\infty}^n\}
=1-\Pr\{X_{n+1}=0\,|\,X_{-\infty}^n=x_{-\infty}^n\}
=\theta_{s(x_{-\infty}^n)}.$$
It is clear that the 
process just defined could be thought of simply 
as a $D$-th order Markov chain,
but this would ignore the important information
contained in $S$: If a suffix string $s\in S$ 
has length $\ell<D$,
then, conditional on any past sequence $x_{-\infty}^n$
which ends in $s$,
the distribution of $X_{n+1}$ only depends on the most 
recent $\ell$ symbols.
Therefore, the suffix set offers an
economical way for describing the transition 
probabilities of $\Xp$, especially 
for chains that can be represented with
a relatively small suffix set.

Suppose that a certain string $x_1^n$
has been generated by a tree process 
of depth no greater than $D$, but with unknown
suffix set $S^*$ and parameter vector
$\thetap^*=(\theta^*_s\;;\;s\in S)$. 
Following classical Bayesian methodology, 
we assign a {\em prior} probability $\pi(S)$ 
on each (complete and proper) suffix set $S$ of
depth $D$ or less, and, given $S$,
we assign a prior probability $\pi(\thetap\,|\,S)$ 
on each parameter vector $\thetap=(\theta_s)$.
A Bayesian approximation to the 
true probability of $x_1^n$
(under $S^*$ and $\thetap^*$)
is the mixture probability,
\be
\hat{P}_{D,\,{\rm mix}}(x_1^n)=\sum_S\pi(S)\int P_{S,\thetaps}(x_1^n)
\pi(\thetap\,|\,S)\,d\thetap,
\label{eq:mix}
\ee
where $P_{S,\thetaps}(x_1^n)$ is the probability
of $x_1^n$ under the distribution of a tree process
with suffix set $S$ and parameter vector $\thetap$.
The expression in (\ref{eq:mix}) is, 
in practice, impossible to compute directly,
since the number of suffix sets of depth $\leq D$
(i.e., the number of terms in the sum) is of order
$2^D$. This is obviously prohibitively 
large for any $D$ beyond 20 or 30.

The CTW algorithm is an efficient procedure
for computing the mixture probability in (\ref{eq:mix}),
for a specific choice of the prior distributions 
$\pi(S),\,\pi(\thetap\,|\,S)$: The prior 
on $S$ is,
$$\pi(S)=2^{-|S|-N(S)+1},$$
where $|S|$ is the number of elements of $S$
and $N(S)$ is the number of string is $S$ with
length strictly smaller than $D$. Given a
suffix set $S$, the prior on $\thetap$ is
the product $(\frac{1}{2},\frac{1}{2})$-Dirichlet distribution,
i.e., under $\pi(\thetap|S)$ the individual $\theta_s$ 
are independent, with each
$\theta_s\sim$ Dirichlet$(\frac{1}{2},\frac{1}{2})$.

The main practical advantage of the CTW algorithm is that
it can actually compute the probability in (\ref{eq:mix})
exactly. In fact, this computation can be performed
sequentially, in linear time in the length of the
string $n$, and using an amount of memory which also
grows linearly with $n$.
This, in particular, makes it possible to consider much longer 
memory lengths $D$ than would be possible with 
the plug-in method.

\bigskip

\noindent{\bf The CTW Entropy Estimator. }
Thus motivated, given a {\em binary} string $x_1^n$, 
we define the {\em CTW entropy estimator $\hat{H}_{n,D,\,{\rm ctw}}$} as,
\be
\hat{H}_{n,D,\,{\rm ctw}} = -\frac{1}{n}\log \hat{P}_{D,\,{\rm mix}}(x_1^n),
\label{eq:CTW}
\ee
where $\hat{P}_{D,\,{\rm mix}}(x_1^n)$ is the mixture probability
in (\ref{eq:mix}) computed by the CTW algorithm. 
[Corresponding proceedures are similarly described in 
\cite{kennel}\cite{london}.]
The justification for this definition comes from 
the discussion leading to equation (\ref{eq:SMBTest}) 
above. Clearly,
if the true probability of $x_1^n$ is $P^*(x_1^n)$, 
the estimator performs well when,
\be
-\frac{1}{n}\log \hat{P}_{D,\,{\rm mix}}(x_1^n)\approx 
-\frac{1}{n}\log P^*(x_1^n).
\label{eq:app}
\ee
In many cases this approximation can be rigorously justified,
and in certain cases it can actually be accurately quantified.

Assume that $x_1^n$ is generated by an
unknown tree process (of depth no greater than $D$)
with suffix set $S^*$.
The main theoretical result of \cite{willems-shtarkov-tjalkens:95}
states that, for {\em any} string $x_1^n$,
of {\em any} finite length $n$,
generated by {\em any} such process,
the difference between the two terms in (\ref{eq:app})
can be uniformly bounded above; from this it easily
follows that,
\be
-\frac{1}{n}\log \hat{P}_{D,\,{\rm mix}}(x_1^n)-\Big[
-\frac{1}{n}\log P^*(x_1^n)\Big]\leq
\frac{|S^*|}{2n}\log n+\frac{3|S^*|+1}{n}.
\label{eq:redundancy}
\ee
This nonasymptotic, quantitative 
bound, easily leads to various properties
of $\hat{H}_{n,D,\,{\rm ctw}}$:

First, (\ref{eq:redundancy})
combined with the Shannon-McMillan-Breiman theorem
(\ref{eq:SMBT}) and the pointwise converse source coding 
theorem \cite{barron:thesis}\cite{kieffer:91},
readily implies that $\hat{H}_{N,D,\,{\rm ctw}}$ is consistent,
that is, it converges to the true entropy rate of
the underlying process, with probability one,
as $n\to\infty$.
Also, Shannon's source coding theorem
\cite{cover:book} implies that
the expected value of $\hat{H}_{n,D,\,{\rm ctw}}$ cannot
be smaller than the true entropy rate $H$;
therefore, taking expectations in (\ref{eq:redundancy})
gives,
\be
0\leq\;[\mbox{bias of }\hat{H}_{n,D,\,{\rm ctw}}]
\;\leq\frac{|S|}{2n}\log n \;+\;O(1).
\label{eq:bias}
\ee
In view of Rissanen's \cite{rissanen:book}
well-known universal lower bound,
(\ref{eq:bias}) shows that 
the bias of the CTW is essentially
as small as can be.
Finally, for the variance, 
if we subtract $H$ from both sides
of (\ref{eq:redundancy}), multiply by $\sqrt{n}$,
and apply the central-limit refinement
to the Shannon-McMillan-Breiman theorem
\cite{yushkevich}\cite{ibragimov:62},
we obtain that 
the standard deviation of
the estimates $\hat{H}_{n,D,\,{\rm ctw}}$ is
$\approx \sigma_X/\sqrt{n}$, where $\sigma_X^2$ is the 
{\em minimal coding variance} of $\Xp$ \cite{kontoyiannis-97}.
This is also optimal, in view of the second-order 
coding theorem of \cite{kontoyiannis-97}.

Therefore, for data $x_1^n$ generated by tree processes,
the bias of the CTW estimator is of 
order $O(\log n/n)$, and its variance is $O(1/n)$.
Compared to the earlier LZ-based estimators,
these bounds suggest much faster convergence, and
are in fact optimal. In particular, the $O(\log n/n)$
bound on the bias indicates that, unlike the LZ-based
estimators, the CTW can give useful results even on
small data sets. 

\bigskip

\noindent
{\bf Extensions. }
An important issue for the 
performance of the CTW entropy estimator, 
especially when used on data 
with potentially long-range dependence,
is the choice of the depth $D$:
While larger values of $D$ give the estimator
a chance to capture longer-term trends,
we then pay a price in the algorithm's 
complexity and in the estimation bias.
This issue will not be discussed further here;
a more detailed discussion of this point along 
with experimental results can be found in 
\cite{gao:thesis}\cite{neuro-isit}.

The CTW algorithm has also been extended beyond finite-memory 
processes \cite{willems:98}. The basic method
is modified to produce an estimated probability
$\hat{P}_\infty(x_1^n)$, without assuming a predetermined
maximal suffix depth $D$. The sequential nature of the
computation remains exactly the same,
leading to a corresponding entropy estimator
defined analogously to the one in (\ref{eq:CTW}),
as
$\hat{H}_{n,\infty,\,{\rm ctw}}=-\frac{1}{n}\log\hat{P}_\infty(x_1^n)$.
Again it is easy to show that 
$\hat{H}_{n,\infty,\,{\rm ctw}}$ is consistent with probability one,
this time for {\em every} stationary and ergodic
(binary) process. The price of this
generalization is that the earlier
estimates for the bias and variance no longer apply,
although they do remain valid is the
data actually come 
from a finite-memory process.
In numerous simulation 
experiments we found that there is 
no significant advantage in using $\hat{H}_{n,D,\,{\rm ctw}}$ 
with a finite depth $D$ over $\hat{H}_{n,\infty,\,{\rm ctw}}$, 
except for the somewhat shorter computation time. 
For that reason, in all the experimental results 
in Sections~\ref{subsec:conver} 
and~\ref{subsec:comp} below, we only report
estimates obtained by 
$\hat{H}_{n,\infty,\,{\rm ctw}}$.

Finally, perhaps the most striking feature of the
CTW algorithm is that it can be modified to compute
the ``best'' suffix set $S$ that can be fitted to 
a given data string, where, 
following standard statistical (Bayesian) practice,
``best'' here means the one which is most 
likely under the posterior distribution.
To be precise, recall the
prior distributions $\pi(S)$ and $\pi(\thetap\,|\,S)$
on suffix sets $S$ and on parameter vectors $\thetap$,
respectively. Using Bayes' rule, the {\em posterior}
distribution on suffix sets $S$ can be expressed,
$$\Pr\{S\,|\,x_1^n\}=
\frac{\int P_{S,\thetaps}(x_1^n)\pi(\thetap\,|\,S)\,d\thetap}
{\hat{P}_{D,\,{\rm mix}}(x_1^n)}\;;$$
the suffix set $\hat{S}=\hat{S}(x_1^n)$ which maximizes
this probability is called the 
{\em Maximum A posteriori Probability}, or MAP,
suffix set.
Although the exact computation of $\hat{S}$
is, typically, prohibitively hard to carry out directly, 
the {\em Context-Tree Maximizing} (CTM) algorithm 
proposed in \cite{maxtree1} is an efficient
procedure (with complexity and memory requirements
essentially identical to the CTW algorithm)
for computing $\hat{S}$. The CTM
algorithm will not be used or discussed further
in this work; see the discussion
in \cite{gao:thesis}\cite{neuro-isit},
where it plays an
important part in the analysis 
of neuronal data.


\section{Results on Simulated Data}
\label{sec:simu}

This section contains the results of an extensive 
simulation study, comparing various aspects of the 
behavior of the different entropy estimators presented 
earlier (the plug-in estimator, the four LZ-based 
estimators, and the CTW estimator), applied 
to different kinds of simulated {\em binary} data. 
Motivated, in part, by applications 
in neuroscience, we also introduce a new method, the 
{\em renewal entropy estimator.}
Section~\ref{subsec:simutrueentropy}
contains descriptions of the statistical models used to 
generate the data,
along with exact formulas or close approximations for
their entropy rates.

Sections~\ref{subsec:conver} and~\ref{subsec:comp}
contain the actual simulation results.

\subsection{Statistical Models and Their Entropy Rates}
\label{subsec:simutrueentropy}

\subsubsection{I.I.D. (or ``homogeneous Poisson'') Data}
The simplest model of a binary random process $\Xp$ is 
as a sequence of independent and identically distributed
(i.i.d.) random variables $\{X_n\}$, 
where the the $X_i$ are independent and all have the same
distribution. This process has no memory, and in the 
neuroscience literature it is often 
referred to as a ``homogeneous Poisson process.'' 
The distribution of each $X_i$
is described by a parameter $p\in[0,1]$, so that
$\Pr\{X_i=1\}=p$ and $\Pr\{X_i=0\}=1-p$. If
$\{X_n\}$ were to represent a spike train, then 
$p=E(X_i)$ would be its average firing rate.

The entropy rate of this process is simply,
$$H = H(X_1) = -p\log p -(1-p)\log(1-p).$$

\subsubsection{Markov Chains}
An $\ell$-th order (homogeneous) Markov chain 
with (finite) alphabet $A$
is a process $\Xp=\{X_n\}$ with the property that,
$$\Pr\{X_{n}=x_n\,|\, X_1^{n-1}=x_1^{n-1}\}
=\Pr\{X_{n}=x_n\,|\, X_{n-\ell}^{n-1}=x_{n-\ell}^{n-1}\}
=P_{x_{n-\ell}^{n-1},x_n},$$
for all $x_1^n\in A^n$,
where $P=\big(P_{x_1^{\ell},x_{\ell+1}}\big)$
is the transition matrix of the chain.
This formalizes the idea
that the memory of the process has length
$\ell$: The probability of each new symbol depends
only on the most recent $\ell$ symbols, and it is
conditionally independent of the more distant
past. The distribution of this process is described
by the initial distribution of its first $\ell$ symbols,
$(\pi(x_1^\ell))$, and the transition matrix $P$.
The entropy rate of an ergodic (i.e., irreducible and aperiodic) 
$\ell$-th order Markov chain $\Xp$ is given by,
\ben
H 
  = -\sum_{x_1^{\ell}} \pi^*(x_1^{\ell}) 
	\sum_{x_{\ell+1}} P_{x_1^{\ell},x_{\ell+1}} \log P_{x_1^{\ell}, 
		x_{\ell+1}},
\een
where $\pi^*$ is the unique stationary distribution of the chain.

\subsubsection{Hidden Markov Models}
\label{subsubsec:HMM}
Next we consider a class of binary processes $\Xp=\{X_n\}$,
called hidden Markov models (HMMs) or hidden
Markov processes, which typically have infinite
memory. 
For the purposes of this discussion, 
a binary HMM can be defined as follows.
Suppose $\Yp=\{Y_n\}$ is a first-order, ergodic Markov chain,
which is stationary,
i.e., its initial distribution $\pi$ is the same
as its stationary distribution $\pi^*$.
Let the alphabet of $\Yp$ be an arbitrary finite set $A$,
write $P=(P_{yy'})$ for its transition matrix,
and let $Q=(Q_{yx}\;;\;y\in A,\,x\in\{0,1\})$ 
be a different transition matrix, from $A$ to $\{0,1\}$.
Then, for each $n$, given 
$\{Y_i\}$ and the previous values $x_1^{n-1}$ of $X_1^{n-1}$, 
the distribution of the random variable $X_n$ 
is,
$$
\Pr\{X_n = x \,|\, Y_n = y\} = Q_{yx},
$$
independently of the remaining $\{Y_i\}$ and of $X_1^{n-1}$.
The resulting process $\Xp=\{X_n\}$ is a binary HMM.

The consideration of HMMs here is partly motivated
by the desire to simulate spike trains with slowly
varying rates, as in the case of real neuronal firing.
To illustrate, consider the following (somewhat 
oversimplified) description of a model that will
be used in the simulation examples below.
Let the Markov chain $\Yp=\{Y_n\}$ represent the 
process which modulates the firing rate of the 
binary ``spike train'' $\Xp$,
so that $\Yp$ takes a finite number of values,
$A=\{r_1,r_2,\ldots,r_\alpha\}$, with each $r_i\in(0,1)$.
These values correspond to $\alpha$ different firing regimes, 
so that, e.g., $Y_n=r_1$ means that the average firing rate 
at that instant is $r_1$ spikes-per-time-unit.
To ensure that the firing rate varies slowly,
define, for every $y\in A$, the transition
probability that $Y_n$ remains in the same state to be
$\Pr\{Y_n=r\,|\,Y_{n-1}=r\}=1-\epsilon$, 
for some small $\epsilon>0$.
Then, conditional 
on $\{Y_n\}$, the distribution of each $X_n$ 
is given by $\Pr\{X_n=1\,|\,Y_n=r\}=r=1-\Pr\{X_n=0\,|\,Y_n=y\}.$

In general, an HMM defined as above is
stationary, ergodic and typically has infinite memory --
it is {\em not} a $\ell$-th order Markov chain 
for any $\ell$; see \cite{ephraim-merhav:02}
and the references therein for details.
Moreover, there is no closed-form expression for the
entropy rate of a general HMM,
but, as outlined below, it is fairly easy to 
obtain an accurate approximation when the
distribution of the HMM is known a priori,
via the Shannon-McMillan-Breiman
theorem (\ref{eq:SMBT}). That is,
the value of the entropy rate $H=H(\Xp)$ can be 
estimated accurately as long as it is possible
to get a close approximation for the 
probability $p_n(X_1^n)$ of a long random sample 
$X_1^n$. This calculation is, in principle, hard 
to perform, since it requires the computation
of an average over all
possible state sequences 
$y_1^n$, and their number grows 
exponentially with $n$.
As it turns out, adapting an idea similar to the
usual dynamic programming algorithm used for HMM
state estimation, the required probability
$p_n(X_1^n)$ 
can actually be computed very efficiently;
similar techniques appear in various places
in the literature, e.g.,
\cite{ephraim-merhav:02}\cite{spa-hmm}.
Here we adopt the following method,
developed independently in \cite{gao:thesis}.

First, generate and fix a long random realization 
$x_1^n$ of the HMM $\Xp$, and
define the matrices $M^{(k)}$ by,
$$
M_{yy'}^{(k)} = P_{yy'} Q_{y'x_k}, \;\; y,y' \in A, 
\;\;\;\; k=2,3,\ldots,n,
$$
and the row vector $b=(b_y)$,
$$
b_y = \pi(y) Q_{yx_1},
\;\;\;\; y\in A.
$$
The following proposition says that 
the probability of an arbitrary $x_1^n$ can 
be obtained in $n$
matrix multiplications, involving matrices of dimension
$|A|\times|A|$. For moderate alphabet sizes, this can
be easily carried out even for large $n$, e.g., on the
order of $10^6.$ Moreover, as the HMM process $\Xp$
inherits the strong mixing properties of the underlying
Markov chain $\Yp$, Ibragimov's central-limit refinement
\cite{ibragimov:62} to the Shannon-McMillan-Breiman theorem 
suggests that the variance of the estimates obtained
should decay at the rapid rate of $1/n$.
Therefore, in practice, 
it should be possible
to efficiently obtain a reliable, stable approximation
for the entropy rate, a prediction we
have repeatedly verified through 
simulation.\footnote{To avoid confusion note that,
although this method gives a very accurate estimate
of the entropy rate of an HMM, in order to carry it 
out it is necessary to know {\em in advance} 
the distributions of both the HMM $\Xp$ {\em and}
of the unobservable process $\Yp$.
}

\begin{proposition} 
\label{prop1}
Under the above assumptions,
the probability of an arbitrary $x_1^n\in\{0,1\}^n$
produced by the HMM $\Xp$ can be expressed as,
$$
p_n(x_1^n) = b\Big [\prod_{k=2}^n M^{(k)}\Big]{\bf 1},
$$
where ${\bf 1}$ is the column vector of $|A|$ 1s.
\end{proposition}


\newpage

\noindent
{\bf Proof:} 
Let $y_1^n\in A^n$ denote any specific realization of the 
hidden Markov process $Y_1^n$. We have,
\begin{eqnarray}
p_n(x_1^n)\;=\;\Pr\{X_1^n=x_1^n\}
&=& 
	\sum_{y_1^n\in A^n} 
	\Pr\{Y_1^n=y_1^n\}
	\Pr\{X_1^n=x_1^n\,|\,Y_1^n=y_1^n\}
	\nonumber \\
&=& 
	\sum_{y_1^n \in A^n}
	\pi(y_1)Q_{y_1x_1}
	\prod_{k=2}^n P_{y_{k-1},y_k} Q_{y_kx_k} 
	\nonumber \\
&=& 
	\sum_{y_n \in A}
	\sum_{y_1 \in A}
	b_{y_1}
	\sum_{y_2^{n-1} \in A^{n-2}}
	\prod_{k=2}^n M^{(k)}_{y_{k-1}y_k} 
	\nonumber \\
&=& 
	\sum_{y_n\in A} \Big(b\Big[\prod_{k=2}^n M^{(k)}
		\Big]\Big)_{y_n} \nonumber \\
&=&  
	b \Big[\prod_{k=2}^n M^{(k)}\Big] {\bf 1}. \nonumber
\end{eqnarray}
\hfill $\Box$

\subsubsection{Renewal Processes}
\label{subsubsec:renewal}
A common alternative mathematical description for 
the distribution of a binary string 
is via the distribution of the
time intervals between successive 1s, or,
in the case of a binary spike train, the 
{\em interspike intervals} (ISIs) as they 
are often called in the neuroscience literature
\cite{Stevens:96}\cite{Bhumbra:04}. 
Specifically, let $\{t_i\}$ denote the sequence of times 
$t$ when $X_t=1$, and let $\{Y_i=t_{i+1}-t_i\}$ be the 
sequence of ``interarrival times'' or ISIs
of $\Xp=\{X_n\}$. Instead of defining the
joint distributions of blocks $X_1^n$ of random
variables from $\Xp$, its distribution can be
specified by that of the process $\Yp=\{Y_i\}$.
For example, if the $Y_i$ are i.i.d.\
random variables with geometric distribution with 
parameter $p\in[0,1]$, then $\Xp$ is itself a (binary)
i.i.d.\ process with parameter $p$.

More generally, a {\em renewal} process $\Xp$
is defined in terms of an arbitrary i.i.d.\
ISI process $\Yp$, with each $Y_i$ having
a common discrete distribution 
$P=(p_j\;;\;j=1,2,\ldots)$.\footnote{The consideration 
	of renewal processes in partly motivated by
	the fact that, as discussed in 
	\cite{neuro-isit}\cite{gao:thesis},
	the main feature of the distribution of
	real neuronal data that gets captured by the 
	CTW algorithm (and by the MAP suffix set 
	it produces) is an empirical estimate of 
	the distribution of the underlying ISI process.
	Also, simulations suggest that
	renewal processes produce firing patterns similar
	to those observed in real neurons firing, 
	indicating that the corresponding entropy estimation
	results are more biologically relevant.}

Recall \cite{papangelou:78} that 
the entropy rates $H(\Xp)$ of $\Xp$ and
$H(\Yp)$ of $\Yp$ are related by,
$$
H(\Xp) = \lambda H(\Yp)=-\lambda\sum_{j=1}^\infty p_j\log p_j,
$$
where $\lambda=E(X_1)=1/E(Y_1)$.
This simple relation motivates the consideration
of a different estimator for the entropy rate 
of a renewal process. Given binary data $x_1^n$
of length $n$, calculate the corresponding
sequence of ISIs $y_1^m$, and define,
\[
\hat{H}_{n,{\rm renewal}} = -\hat{\lambda} \sum_j \hat{q}_j \log \hat{q}_j,
\]
where $\hat{Q}=(\hat{q}_j)$ is the empirical distribution 
of the ISIs $y_1^m$, and $\hat{\lambda}$ is the empirical rate
of $\Xp$, i.e., the proportion of 1s in $x_1^n$.
We call this the {\em renewal entropy rate estimator.}

When the data are indeed generated by a renewal process,
the law of large numbers guarantees that $\hat{H}_{n,{\rm renewal}}$ 
will converge to $H(\Xp),$ with probability one, 
as $n\to\infty$. Moreover, under rather 
weak regularity conditions on the distribution of the 
ISIs, the central limit theorem implies that the variance
of these estimates decays like $O(1/n)$.
But the results of the renewal entropy estimator remain
meaningful even if $\Xp$ is not a renewal process. 
For example, if the corresponding ISI process
$\Yp$ is stationary and ergodic but not i.i.d., 
then the renewal entropy estimator will 
converge to the value $\lambda H(Y_1)$, whereas the true 
entropy rate in this case is the (smaller) quantity,
$$H(\Yp)=\lim_{i\to\infty} \lambda H(Y_i\,|\,Y_{i-1},\ldots,Y_2,Y_1).$$
Therefore, 
the renewal entropy estimator can be employed
to test for the presence or absence 
of renewal structure in particular data sets,
by comparing the value of 
$\hat{H}_{n,{\rm renewal}}$ with that
of other estimators; see \cite{neuro-isit}\cite{gao:thesis}
for a detailed such study.

\subsection{Bias and Variance of the CTW and the LZ-based Estimators}
\label{subsec:conver}

In order to compare the empirical bias and variance 
of the four LZ-based estimators and the CTW estimator 
with the corresponding theoretical predictions 
of Sections~\ref{subsec:biasvarlz} and~\ref{subsec:ctwtheory},
the five estimators are applied to simulated (binary) 
data. [In this as well as in 
	the following section,  we only present results
	for the {\em infinite}-suffix-depth CTW estimator
	$\hat{H}_{n,\infty,\,{\rm ctw}}$;
	recall the discussion
	at the very end Section~\ref{subsec:ctwtheory}.]
The simulated data were generated from four different 
processes: An i.i.d.\ (or ``homogeneous Poisson'') 
process, and three different Markov chains
with orders $\ell=1,2,$ and 10.
For each parameter setting of each model,
50 independent realizations were generated
and the bias of each method was estimated
by subtracting the true entropy rate
from the empirical mean of the individual 
estimates. Similarly, the standard error 
was estimated by the empirical standard 
deviation of these results.

\begin{figure}[ht]
\centerline{
\psfig{figure=            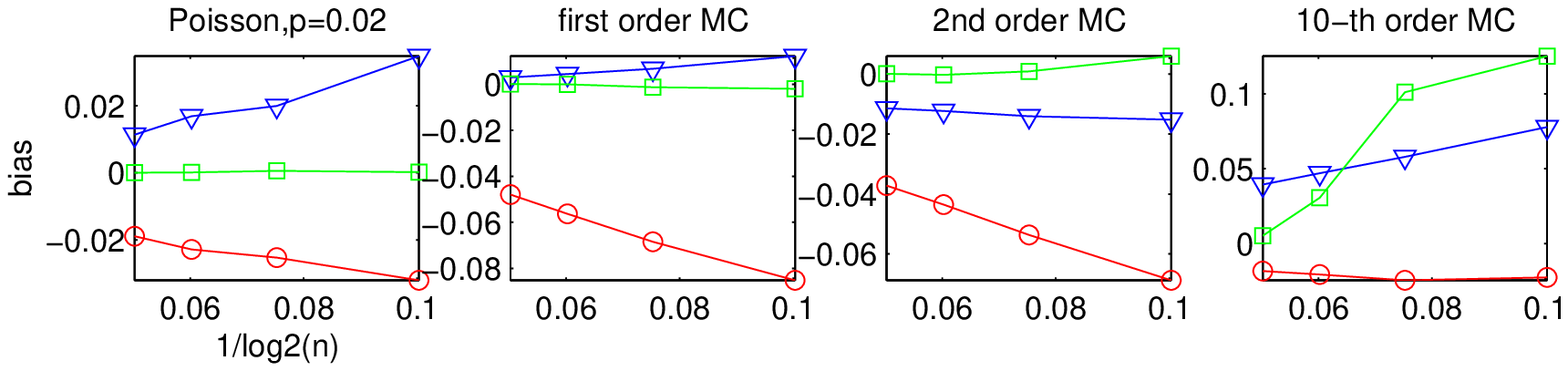,width=5.5in}
}
\centerline{
\hspace{0.2in}
\psfig{figure=            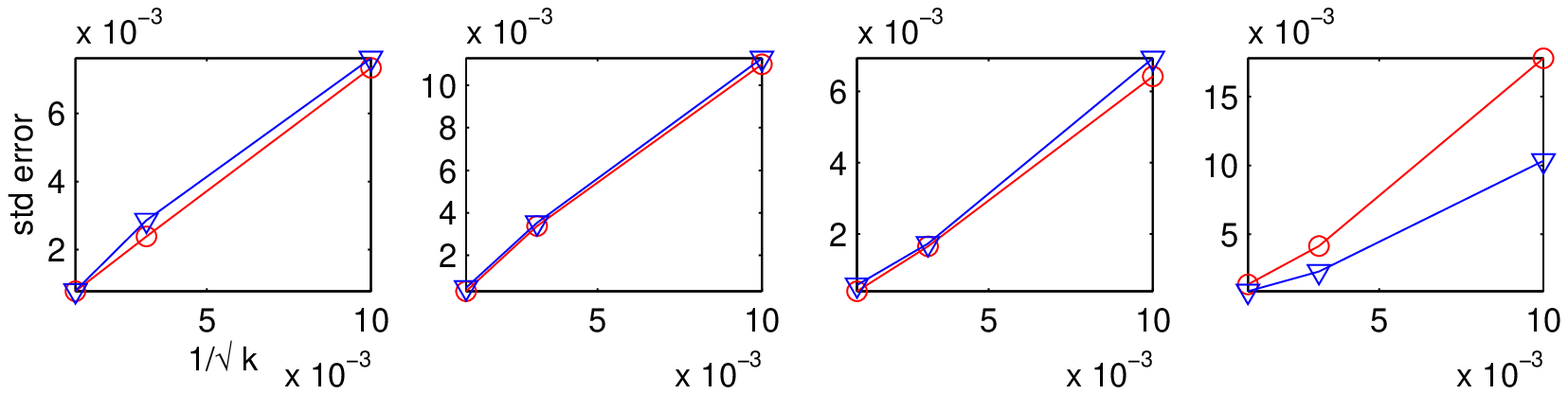,width=5.3in}
}
\caption{Results obtained by $\hat{H}_{n,k}$ (shown as red 
lines with circles), by $\tilde{H}_{n,k}$ (blue lines with 
triangles), and by the CTW estimator $\hat{H}_{N,\infty,\,{\rm ctw}}$
(green lines with squares),
applied to data from four different processes.  
For $\hat{H}_{n,k}$ and $\tilde{H}_{n,k}$,
the first row shows the bias plotted against
$1/\log n$, for window lengths $n=10^3,10^4,10^5, 10^6$,
and with a fixed number of matches $k=10^6$;
the second row shows the standard error 
plotted against $1/\sqrt{k}$, where $k=10^4,10^5,10^6$,  
with fixed $n=10^6$. 
For $\hat{H}_{N,\infty,{\rm ctw}}$, 
the first row shows the bias when applied to
data of the same total length $N=n+2k$; for its 
standard error 
see Figure~\ref{fg:conv_incrswin}.
The true values of the entropy rates of the four processes are
$H=0.1414$, $H=0.4971$, $H=0.7479$ and $H=0.6946$, respectively.}
\label{fg:conv_fixwin}
\end{figure}

\begin{figure}[ht]
\centerline{
\psfig{figure=            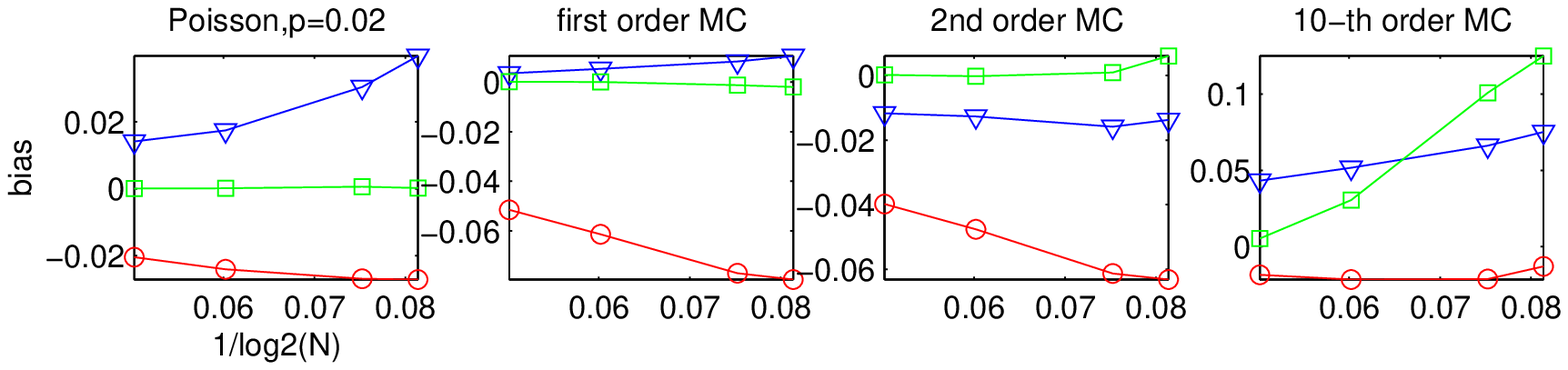,width=5.5in}
}
\centerline{
\hspace{0.1in}
\psfig{figure=            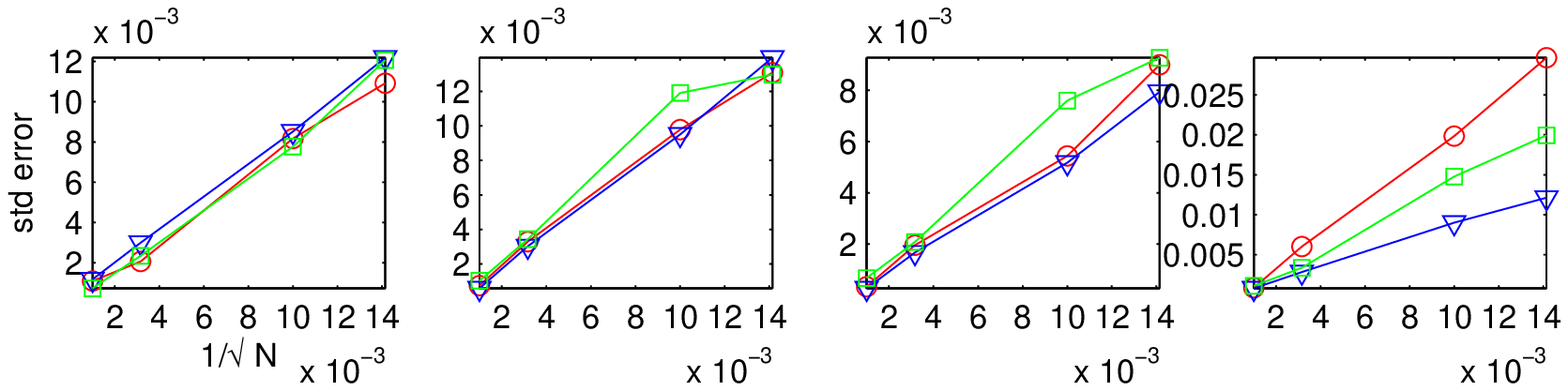,width=5.3in}
}
\caption{Results obtained by $\hat{H}_n$ (shown as red 
lines with circles), by $\tilde{H}_n$ (blue lines with 
triangles), and by the CTW estimator (green lines with squares),
applied to data from four different processes.  
For $\hat{H}_n$ and $\tilde{H}_n$,
the first row shows the bias plotted against
$1/\log n$, for window lengths $n=5000,10^4,10^5, 10^6$,
and the second row shows the standard error 
plotted against $1/\sqrt{n}$.
Similarly, for $\hat{H}_{n,\infty,{\rm ctw}}$, 
the two rows show the bias and standard error
when the estimator is applied to
data of the same total length.
The true values of the entropy rates of 
the four processes are the same as before.}
\label{fg:conv_incrswin}
\end{figure}

Specifically, for $\hat{H}_{n,k}$ and $\tilde{H}_{n,k}$,
first a window size $n$ and a number of matches $k$
were selected, and then 50 independent realizations of 
length $N=2n+k$ were generated. The bias and standard
error are plotted in Figure~\ref{fg:conv_fixwin} against
$O(1/\log n)$ and $O(1/\sqrt{k})$, respectively. The 
approximately linear curves confirm the theoretical 
predictions that the bias and variance decay to zero 
at rates $O(1/\log n)$ and $O(1/k)$, respectively. 
Note that, although $\tilde{H}_{n,k}$
is always larger than $\hat{H}_{n,k}$, there is
no systematic trend regarding which one gives
more accurate estimates.
Also plotted on Figure~\ref{fg:conv_fixwin} is the bias 
of infinite-suffix-depth CTW estimator,
$\hat{H}_{N,\infty,{\rm ctw}}$, applied 
to data with total length
$N=2n+k$. 
[The estimated standard error of the CTW 
estimator is not shown in Figure~\ref{fg:conv_fixwin},
because it cannot be compared to the behavior of the
LZ-based estimators as the number of matches $k$
varies; see Figure~\ref{fg:conv_incrswin}
for estimates of the CTW standard error
on the same set of experiments.]

The results of the CTW are generally much more 
accurate than those of the LZ estimators,
except in the case of the 10th order Markov 
chain with small data size, where
the LZ-based methods seem to get better results.

The values of the bias and 
standard error of $\hat{H}_{n,k}$ and $\tilde{H}_{n,k}$ 
in Figure~\ref{fg:conv_fixwin}  
suggest that, in order to minimize the total
mean squared error (MSE) of the LZ-based estimates, 
the number of matches $k$ should be chosen to be
small relative to the window length $n$,
since the variance clearly appears to decay 
much faster than the square of the bias. 
This is further confirmed by the results shown 
in Table~\ref{tab:simu_tradeoff}, which shows
corresponding estimates for an
i.i.d.\ process with $p=0.25$ and data length $N=10^6$. 
The values of $n$ and $k$ satisfy $n+k=N-2\log N$. 
The ratio $n/k$ ranges from 1 to 10,000. 

\begin{table}[ht]
  \begin{center}
    \begin{tabular}{|r|r|r||c|c|c||c|c|c|}
      \hline
      \multicolumn{3}{|c||}{} & \multicolumn{3}{|c||}{$\hat{H}_{n,k}$} & 
\multicolumn{3}{|c|}{$\tilde{H}_{n,k}$} \\
      \hline
    \multicolumn{1}{|c|}{$n/k$} & \multicolumn{1}{|c|}{$n$} &
\multicolumn{1}{|c||}{$k$} & bias &
std err & MSE & bias & std err & MSE \\
     \hline
1 & 499980 & 499980  & -0.0604 & 0.0010 & 0.1824 & -0.0325 & 0.0009 & 0.0528\\
10 & 909054 & 90906  & -0.0584 & 0.0018 & 0.1705 & -0.0318 & 0.0019 & 0.0507\\
100 & 990059 & 9901  & -0.0578 & 0.0066 & 0.1692 & -0.0315 & 0.0067 & 0.0517\\
1000 & 998961 & 999  & -0.0553 & 0.0200 & 0.1732 & -0.0297 & 0.0210 & 0.0663\\
10000 & 999860 & 100  & -0.0563 & 0.0570 & 0.3209 & -0.0356 & 0.0574 & 0.2280\\
     \hline 
     \end{tabular}
     \medskip
     \caption{Choosing $n$ and $k$ to minimize MSE.  Data are generated 
from an i.i.d.\ process with $p=0.25$, the true entropy
rate is $H=0.8113$, the data length $N=10^6$, and $n+k=N-2\log N$. }
     \label{tab:simu_tradeoff}
  \end{center}
\end{table}

Next, the validity of the bootstrap procedure for the 
standard error of the LZ estimators $\hat{H}_{n,k}$ 
and $\tilde{H}_{n,k}$ is examined; recall the
description in Section~\ref{subsec:biasvarlz}.
For data generated from the same four processes,
the bootstrap estimate of the standard
error is compared to that obtained empirically from 
the sample standard deviation computed from 50 independent
repetitions of the same experiment. Since that the natural
domain of applicability of the bootstrap method is
for large values of $k$, Tables~\ref{tab:simu_bstrp1} 
and~\ref{tab:simu_bstrp2} contain simulation results
for $k=10^5$ and $k=10^6,$ with $n=10^3$. The results clearly
indicate that the bootstrap procedure indeed gives accurate 
estimates of the standard error.

\begin{table}[ht]
  \begin{center}
    \begin{tabular}{|c||c|c|c|c|}
      \hline 
      & \multicolumn{4}{|c|}{$\hat{H}_{n,k}$} \\
      \hline
      & \multicolumn{2}{|c}{$k=10^5$} &
\multicolumn{2}{|c|}{$k=10^6$}\\
        \hline
    case & bootstrap & std dev & bootstrap & std dev\\
     \hline
1 & 0.0018 & 0.0025 & 0.0006 & 0.0009 \\
2 & 0.0025 & 0.0023 & 0.0008 & 0.0009 \\
3 & 0.0015 & 0.0019 & 0.0005 & 0.0007 \\
4 & 0.0061 & 0.0073 & 0.0017 & 0.0023 \\
     \hline 
     \end{tabular}
     \medskip
     \caption{Comparison between the bootstrap standard error
	and the empirical estimate of the standard deviation
	for the four different processes.
 	The window size $n$ is fixed at $10^3$.}
     \label{tab:simu_bstrp1}
  \end{center}
\end{table}

\begin{table}[ht]
  \begin{center}
    \begin{tabular}{|c||c|c|c|c|}
      \hline 
      & \multicolumn{4}{|c|}{$\tilde{H}_{n,k}$} \\
      \hline
      & \multicolumn{2}{|c}{$k=10^5$} &
\multicolumn{2}{|c|}{$k=10^6$}\\
        \hline
    case & bootstrap & std dev & bootstrap & std dev\\
     \hline
1 & 0.0033 & 0.0033 & 0.0011 & 0.0010 \\
2 & 0.0031 & 0.0025 & 0.0010 & 0.0009 \\
3 & 0.0017 & 0.0016 & 0.0006 & 0.0006 \\
4 & 0.0032 & 0.0031 & 0.0009 & 0.0010 \\
     \hline 
     \end{tabular}
     \medskip
     \caption{Comparison between the bootstrap standard error 
	and the empirical estimate of the standard deviation
	for the four different processes.
 The window size $n$ is fixed at $10^3$.}
     \label{tab:simu_bstrp2}
  \end{center}
\end{table}

Corresponding experiments were performed
for $\hat{H}_n$ and $\tilde{H}_n$,
applied to data from the same four types of processes.
Although for these estimators there is little in the
way of rigorous theory that can be used as a guideline
to compute their mean and variance, simple heuristic 
calculations strongly suggest that they should decay like
$O(1/\log n)$ and $O(1/n)$, respectively.
Figure~\ref{fg:conv_incrswin} shows the bias and 
standard error computed empirically as for 
$\hat{H}_{n,k}$ and $\tilde{H}_{n,k}$, and
plotted against $1/\log n$ and 
$1/\sqrt{n}$, respectively. Once again,
the fact that the resulting empirical curves are
approximately linear agrees with these 
predictions.  Observe that 
the bias of $\tilde{H}_n$ can be either
positive or negative; the same behavior
was observed for $\hat{H}_n$ in 
numerous simulation experiments.

The fact that the
standard error converges much faster 
than the bias strongly suggests that,
for all four LZ-based estimators,
it is the {\em bias} that dominates the estimation error,
even in these simple cases of processes with
fairly short memory.

Figure~\ref{fg:conv_incrswin} also shows the bias and standard 
error of the CTW estimator $\hat{H}_{n,\infty,\,{\rm ctw}}$. 
Its bias appears to generally converge 
significantly faster than that of the LZ-based methods,
while its standard error is very close
to that of $\hat{H}_n$ and $\tilde{H}_n$, 
apparently converging to zero at a very 
similar rate. Nevertheless, the results
show that the main source of error of the
CTW is again from the bias.

Finally we note that, in the results shown, as well
as in numerous other simulation runs, we found 
that the two increasing-window LZ estimators
$\hat{H}_n$ and $\tilde{H}_n$ generally performed
better, and always at least as well, as the corresponding
sliding-window estimators 
$\hat{H}_{n,k}$ and $\tilde{H}_{n,k}$. Specifically,
the simulation results show that the optimal choice 
of parameters for $\hat{H}_{n,k}$ and $\tilde{H}_{n,k}$ 
leads to estimates whose bias is very similar
to that of $\hat{H}_n$ and $\tilde{H}_n$,
whereas the increasing-window estimators
have significantly smaller variance.

\subsection{Comparison of the Different Entropy Estimators}
\label{subsec:comp}

This section contains a systematic comparison
of the performance of all the estimators introduced 
so far: The plug-in method, the LZ-based estimators, 
the CTW algorithm, and the renewal entropy estimator.
All the estimators are applied to different 
types of simulated binary data, generated from processes 
with different degrees of dependence and memory.
The main figure of merit adopted here is the {\em ratio} 
$\frac{\sqrt{\rm MSE}}{H}$,
between the square-root of the mean square error (MSE),
and the true entropy rate. The corresponding ratios
of the bias to the entropy and of the standard error
to the entropy are also considered.

Since, as noted earlier, 
the two increasing-window LZ estimators $\hat{H}_n,\,\tilde{H}_n$
generally perform better or at least as well as their sliding-window 
counterparts, $\hat{H}_{n,k},\,\tilde{H}_{n,k}$,
from now on we restrict attention
to $\hat{H}_n$ and $\tilde{H}_n$.

Table~\ref{tab:simu_msecomp1} shows the results
obtained by the plug-in for word-lengths $w=15$ and $w=20$, 
the LZ-estimators $\hat{H}_n$ and $\tilde{H}_n$, and the CTW
estimator $\hat{H}_{n,\infty,\,{\rm ctw}}$,
on data generated from the same four finite-memory 
processes as above: An i.i.d.\
process and three Markov chains of order
$\ell=1,2$ and 10. 
Again we observe
that the main source 
of error is from the bias for all five 
estimators.
The CTW appears to have the smallest bias,
while the standard error varies little among
the different estimators. 
More specifically, the results demonstrate 
that, for short memory processes, the plug-in estimates
are often better than those obtained by the LZ estimators,
whereas for the 10-th order Markov 
chain the plug-in with word-length $w=20$ is much worse 
than $\hat{H}_n$ and $\tilde{H}_n$ because of the 
undersampling problem mentioned earlier.  
The CTW estimator performs uniformly
better than all the other estimators, for both short and 
relatively long memory processes.  Its fast convergence 
rate outperforms the LZ-based estimators, and its ability 
to look much further into the past makes it much more
accurate than the plug-in.

\begin{table}[ht]
  \begin{center}
    \begin{tabular}{|c|c|c|c|c|c|c|}
      \hline
    & &plug-in & plug-in  & &  &\\
{\em Data model} & &$w=15$ & $w=20$ & $\hat{H}_n$ & $\tilde{H}_n$ & CTW\\
\hline
i.i.d.&\% of bias & 0.001 & -0.10 & -14.47 & 9.98 & 0.04\\
&\% of stderr & 0.57 & 0.51 & 0.77 & 0.83 & 0.51\\
&\% of $\sqrt{MSE}$  & {\bf 0.57} & {\bf 0.52} & {\bf 14.49} & {\bf 10.01} & {\bf 0.52}\\
\hline
1st order MC&\% of bias & -0.11 & -0.78 & -10.38 & 0.70 & 0.02\\
&\% of stderr & 0.22 & 0.16 & 0.15 & 0.12 & 0.21\\
&\% of $\sqrt{MSE}$  & {\bf 0.25} & {\bf 0.80} & {\bf 10.38} & {\bf 0.71} & {\bf 0.21}\\
\hline
2nd order MC&\% of bias  & 4.16 & 1.72 & -5.32 & -1.56 & 0.02\\
&\% of stderr & 0.08 & 0.08 & 0.04 & 0.04 & 0.09\\
&\% of $\sqrt{MSE}$ & {\bf 4.16} & {\bf 1.73} & {\bf 5.32} & {\bf 1.56} & {\bf 0.09}\\
\hline
10th order MC&\% of bias & 16.04 & 10.03 & -2.66 & 6.23 & 0.77\\
&\% of stderr & 0.19 & 0.14 & 0.13 & 0.12 & 0.16\\
&\% of $\sqrt{MSE}$ & {\bf 16.04} & {\bf 10.04} & {\bf 2.66} & {\bf 6.24} & {\bf 0.78}\\
\hline
    \end{tabular}
    \medskip
    \caption{Comparison of the ratios between the bias, the standard error 
	and the square-root of the MSE over the true entropy rate.	
	Five estimators are used on four different types 
	of data. All estimators are applied to data of the same
	length $n=10^6$. Results are shown as {\bf percentages},
	that is, all ratios are multiplied by 100.}
    \label{tab:simu_msecomp1}
  \end{center}
\end{table}

Table~\ref{tab:simu_msecomp2} shows corresponding results 
for data generated by two different binary HMMs; 
recall the relevant description from Section~\ref{subsubsec:HMM}.
The first one has three (hidden) states, $A=\{r_1,r_2,r_3\}$,
where $r_1=0.005$, $r_2=0.02$, $r_3=0.05$. The transition
matrix of $\Yp$ has 
$\Pr\{Y_{n+1}=r\,|\,Y_n=r\}=1-\epsilon$,
and 
$\Pr\{Y_{n+1}=r'\,|\,Y_n=r\}=\epsilon/2$,
for any $r$ and $r'\neq r$, where
$\epsilon=0.001$.
Given the sequence $\Yp=\{Y_n\}$ of hidden states,
the observations $\Xp=\{X_n\}$ are conditionally independent 
samples, where each $X_n$ is a binary random variable
with $\PR\{X_n=1|Y_n=r\}=1-\PR\{X_n=0|Y_n=r\}=r$.
In the second example, the
hidden process $\Yp=\{Y_n\}$ takes values in the set $A$ 
of 50 real numbers evenly spaced between 0.001 and 0.1.
The evolution of the process $\Yp=\{Y_n\}$ is that of a
nearest-neighbor random walk; $Y_n$ stays constant
with probability $(1-\epsilon)$ and it moves to either
one of its two neighbors with probability $\epsilon/2$,
with $\epsilon=0.02$. The conditional distribution
of $\Xp$ given $\Yp$ is the same as before.

\begin{table}[ht]
  \begin{center}
    \begin{tabular}{|c|c|c|c|c|c|c|}
      \hline
    & &plug-in & plug-in  & &  &\\
{\em HMM} & &$w=15$ & $w=20$ & $\hat{H}_n$ & $\tilde{H}_n$ & CTW\\
\hline
3 states &\% of bias  & 4.05 & 3.69 & -43.41 & 11.46 & 2.51\\
&\% of stderr & 2.43 & 2.43 & 1.79 & 2.64 & 2.41\\
&\% of $\sqrt{MSE}$ & {\bf 4.74} & {\bf 4.43} & {\bf 43.47} & {\bf 11.75} & {\bf 3.50}\\
\hline
50 states&\% of bias & 2.98 & 2.53 & -35.62 & 5.39 & 2.31\\
&\% of stderr & 3.26 & 3.25 & 3.15 & 3.35 & 3.26\\
&\% of $\sqrt{MSE}$  & {\bf 4.43} & {\bf 4.12} & {\bf 35.76} & {\bf 6.33} & {\bf 4.00}\\
\hline
    \end{tabular}
    \medskip
    \caption{Comparison of the ratios between the bias, the standard error 
	and the square-root of the MSE over the true entropy rate.	
	Five estimators are used on data generated from
	two different HMMs.
	All estimators are applied to data of the same
	length $n=10^6$. Results are shown as {\bf percentages},
	that is, all ratios are multiplied by 100.}
    \label{tab:simu_msecomp2}
  \end{center}
\end{table}

In both cases, the ``true'' entropy of the underlying
HMM was computed using the
approximation procedure described earlier.
In the first example,
10 independent realizations of the HMM were used 
according to the formula of Proposition~\ref{prop1}, and the 
average of the resulting estimates was taken to be 
the true entropy rate in the calculations
of the bias and standard error in 
Table~\ref{tab:simu_msecomp2}. The same
procedure was applied to three independent
realizations of the second example.

The results on HMM data are quite similar to those 
obtained on processes with short memory shown 
in Table~\ref{tab:simu_msecomp1}.
The LZ-based estimators have heavy biases,
whereas both the plug-in and the CTW method 
give very accurate estimates. This is probably
due to the fact that, although HMMs in general
have infinite memory, the memory of an HMM with 
a finite number of states decays exponentially,
therefore only the short-range statistical 
dependence in the data is significant.

To simulate binary data strings with potentially
longer memory (and with characteristics that are 
generally closer to real neuronal spike trains), 
we turn to renewal processes 
$\Xp=\{X_n\}$ with ISIs $\Yp=\{Y_i\}$ that are 
distributed according to a mixture
of discretized Gamma distributions;
recall the description of 
Section~\ref{subsubsec:renewal}.
Here the $Y_i$ are taken to be
i.i.d.\ with distribution $P=(p_j)$ given by,
$$p_j=\mu f_1(j)+(1-\mu) f_2(j),\;\;\;\;j=1,2,3,\ldots,$$
where $\mu\in(0,1)$ is the mixing proportion
and each $f_i$ is a (discretized) Gamma density
with parameters $(\alpha_i,\beta_i)$, 
respectively.

\begin{figure}[ht]
\centerline{
\psfig{figure=            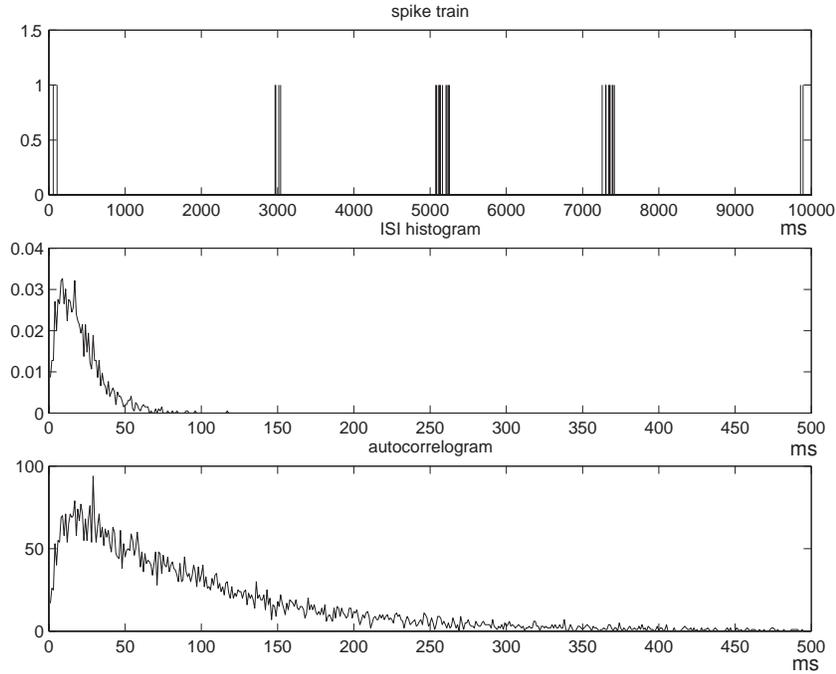,width=4.4in}
}
\caption{
Simulated binary renewal process with 
characteristics similar to those of a bursty neuron.
The ISI distribution is given by a mixture of discrete
Gammas, $\frac{9}{10} f_{(2,10)}+\frac{1}{10} f_{(50,50)}$.
The ``rate'' of the process is $E(X_1)=0.00398.$
The first plot shows the first 10000 simulated bits
of $\Xp$; the second plot shows the empirical 
ISI distribution; the last plot shows the autocorrelogram
of $\Xp$ for values of the lag between 0 and 500.
}
\label{fg:renewalburst}
\end{figure}

If the ISI distribution $P$ was simply Gamma$(\alpha,\beta)$,
then the rate $E(X_1)$ of the binary process $\Xp$
would be the reciprocal of $E(Y_1)=\alpha\beta$;
therefore, taking a mixture of two Gamma densities,
one with $\alpha\beta$ small and one with $\alpha\beta$
large, gives an approximate model of a ``bursty'' neuron,
that is, a neuron which sometimes fires several times
in rapid succession, and sometimes does not fire for
a long period.
Figure~\ref{fg:renewalburst} shows the result 
of such a simulated process $\Xp$. 
The empirical ISI density resembles a real neuron's ISI
distribution, and the peak near the zero lag in the 
autocorrelogram of $\{X_n\}$ indicates the bursty
behavior.

Table~\ref{tab:simu_msecomp3} shows the
results of the estimates 
obtained by the renewal entropy
estimator, the plug-in with word-length $w=20$, 
the two increasing-window LZ-based estimators, 
and the CTW method, applied to data generated 
by a binary renewal process. The
ISI distribution $P$ is a mixture of
two (discrete) Gamma densities $f_1$ and
$f_2$, where $f_1$ has fixed
parameters $(\alpha_1,\beta_1)=(2,10)$
that represent the bursting regime,
and $f_2$ takes three different sets 
of (much larger) parameters $(\alpha_2,\beta_2)$, 
representing low frequency firing.

\begin{table}[ht]
  \begin{center}
    \begin{tabular}{|c|c|c|c|c|c|c|}
      \hline
    &  & renewal& plug-in  & &   & \\
$(\alpha_2,\beta_2)$ & & entropy & $w=20$ & $\hat{H}_n$ & $\tilde{H}_n$ & CTW\\
\hline
(10,20)&\% of bias & -0.06 & 6.09 & -20.98 & 21.84 & 1.66\\
&\% of stderr & 0.73 & 0.77 & 0.66 & 0.95 & 0.72\\
&\% of $\sqrt{MSE}$ & {\bf 0.74} & {\bf 6.14} & {\bf 20.99} & {\bf 21.86} & {\bf 1.81}\\
\hline
(50,20)&\% of bias & -1.53 & 25.90 & -30.38 & 81.40 & 7.64\\
&\% of stderr & 2.37 & 3.03 & 0.79 & 2.67 & 2.38\\
&\% of $\sqrt{MSE}$ & {\bf 2.82} & {\bf 26.08} & {\bf 30.39} & {\bf 81.45} & {\bf 8.00}\\
\hline
(50,50)&\% of bias & -5.32 & 34.35 & -50.64 & 85.61 & 3.58\\
&\% of stderr & 2.36 & 3.12 & 0.69 & 2.67 & 2.42\\
&\% of $\sqrt{MSE} $& {\bf 5.82} & {\bf 34.49} & {\bf 50.65} & {\bf 85.65} & {\bf 4.32}\\
\hline
    \end{tabular}
    \caption{Comparison of the ratios between the bias, the standard error 
	and the square-root of the MSE over the true entropy rate.	
	Five estimators are used on three different data sets
	generated by a renewal process whose ISI distribution $P$
	is a mixture of Gammas, 
	$P=\mu f_{(2,10)}+(1-\mu) f_{(\alpha_2,\beta_2)}$.
	The mixing parameter $\mu=0.8$ in the first two cases,
        and $\mu=0.9$ in the last case.
	All estimators are applied to data of the same total
	length $n=10^6$. Results are shown as {\bf percentages},
	that is, all ratios are multiplied by 100.
}
    \label{tab:simu_msecomp3}
  \end{center}
\end{table}

The CTW and renewal estimator consistently outperform the other
methods, and, on the other extreme, the LZ-based estimators
have high biases and give relatively poor results.
The plug-in estimator only considers what happens in 20-bit-long 
windows, and therefore it misses all the data features
beyond this range. Especially in the case of large parameters 
$(\alpha_2,\beta_2)$ where the ISI distribution has heavier
tails, the structure and the dependence in the data becomes 
significantly longer in range. Also, in that regime, the 
resulting number of ISI data points $y_i$ is much smaller, 
and therefore, the empirical ISI distribution is less accurate; 
as a result, the renewal entropy estimator 
also becomes less accurate. The situation is similar for the
CTW. As shown in \cite{gao:thesis}\cite{neuro-isit},
the CTW method also essentially approximates the empirical 
ISI distribution (although it does so in a different, 
more efficient way than the renewal entropy estimator),
and for larger values of $(\alpha_2,\beta_2)$ its estimates
become less accurate, in a way analogous to the renewal entropy 
estimator.

\newpage

\section{Summary and Concluding Remarks}
\label{sec:concl}

A systematic and extensive comparison between several of the
most commonly used and effective entropy estimators for
binary time series was presented.
Those were the plug-in or ``maximum likelihood'' estimator,
four different LZ-based estimators, the CTW method, and
the renewal entropy estimator.

\medskip

\noindent
{\em Methodology.} Three new entropy estimators
were introduced; two new LZ-based estimators,
and the renewal entropy estimator, which is
tailored to data generated by a binary renewal 
process. A bootstrap procedure, similar to the one
employed in \cite{Suzuki:99}, was described,
for evaluating the standard error of the two
sliding-window LZ estimators $\hat{H}_{n,k}$
and $\tilde{H}_{n,k}$. Also, for these two estimators, 
a practical rule of thumb was heuristically derived 
for selecting the values of the parameters $n$ and $k$ 
in practice.

\medskip

\noindent
{\em Theory.} It was shown 
(Theorem~\ref{thm1}) that the two new LZ-based
estimators $\tilde{H}_{n,k}$ and $\tilde{H}_n$
are {\em universally} consistent, that is,
they converge to the entropy rate for every 
finite-valued, stationary and ergodic process.
Unlike the corresponding
estimators $\hat{H}_{n,k}$ and $\hat{H}_n$
of \cite{kasw}, no additional conditions
are required. An effective method was derived 
(Proposition~\ref{prop1}) for the accurate approximation 
of the entropy rate of a finite-state HMM with
known distribution. Heuristic calculations were presented
and approximate formulas derived for evaluating 
the bias and the standard error of each estimator.

\medskip

\noindent
{\em Simulation.} Several general conclusions 
can be drawn from the simulation experiments 
conducted.
\begin{MyEnumerate}
\item
For all estimators considered, the main source 
of error is the bias. 
\item
Among all the different estimators,
the CTW method is the most effective;
it was repeatedly and consistently seen to
provide the most accurate and reliable results. 
\item
No significant benefit is derived 
from using the finite-context-depth version
of the CTW. 
\item
Among the four LZ-based estimators, 
the two most efficient ones are those with increasing 
window sizes, $\hat{H}_n,$ $\tilde{H_n}$. 
No systematic trend was observed regarding which 
one of the two is more accurate. 
\item
Interestingly (and somewhat surprisingly),
in many of our experiments
the performance of the LZ-based estimators was
quite similar to that of the plug-in method.
\item
The main drawback of the plug-in method is 
its computational inefficiency; with small 
word-lengths it fails to detect longer-range structure 
in the data, and with longer word-lengths the empirical 
distribution is severely undersampled, leading to 
large biases. 
\item
The renewal entropy estimator,
which is only consistent for 
data generated by a renewal process,
suffers a drawback similar 
(although perhaps less severe) 
to the plug-in.
\end{MyEnumerate}

\bigskip

In closing we note that much of the work reported here 
was done as part of the first author's Ph.D.\ thesis
\cite{gao:thesis}. Several new estimators and 
results have appeared in the literature since then, 
perhaps most notably in \cite{C-K-Verdu:04}.
There, a different entropy estimator is introduced
based on the Burrows-Wheeler transform (BWT), it is
shown to be consistent for all stationary and ergodic
processes, and bounds on its convergence rate are
obtained. Moreover, simulation experiments on
binary data indicate that it significantly 
outperforms the plug-in estimator as well as
a modified version of the LZ-based estimator 
$\hat{H}_n$ of \cite{kasw}.
In view of the present results,
interesting further work would include a detailed
comparison of the BWT estimator of \cite{C-K-Verdu:04}
with the CTW method.

%

\subsubsection*{Acknowledgments}
Y.\ Gao was supported by the Burroughs Welcome fund.  
I.\ Kontoyiannis was supported in part
by a Sloan Foundation Research Fellowship and by
NSF grant \#0073378-CCR.
E.\ Bienenstock was supported by NSF-ITR 
Grant \#0113679 and NINDS Contract N01-NS-9-2322.

\bibliographystyle{plain}


\end{document}